\documentclass{gSTA2e}
\usepackage{amsmath}
\usepackage{subfigure}
\usepackage{amssymb}
\usepackage{mathptmx}      % use Times fonts if available on your TeX system
\usepackage[margin=10pt,font=footnotesize,format=hang,justification=justified,labelfont=bf,labelsep=space]{caption}
\newtheorem{Remark}[theorem]{\bf Remark}
\newtheorem{Algorithm}[theorem]{\bf Algorithm}

\RequirePackage{natbib}
\usepackage{url}
\RequirePackage{color}

\newcommand{\href}[2]{{#2}}
\RequirePackage{ifthen}

\newcommand{\rfia}[1]{\makebox[\parindent][l]{%
                     \makebox[0em][r]{\rm(}\sf#1\rm)}}
\newcounter{ABCcB}
\newcommand{\theABCcC}{\alph{ABCcB}}

\newcommand{\Ew}{\mathop{\rm {{}E{}}}\nolimits} % mathop erkennt nur (
\newcommand{\ve}{\varepsilon}
\newcommand{\IF}{\mathop{\rm IF}\nolimits}
\newcommand{\Ts}{\textstyle}

\newcommand{\SSs}{\scriptscriptstyle}
\newcommand{\ssr}{\rm\scriptscriptstyle}

\newcommand{\Lo}{\mathop{\rm {{}o{}}}\nolimits}
\newcommand{\LO}{\mathop{\rm {{}O{}}}\nolimits}
\newcommand{\asVar}{\mathop{\rm{} asVar{}}\nolimits}
\newcommand{\asMSE}{\mathop{\rm{} asMSE{}}\nolimits}
\newcommand{\asBias}{\mathop{\rm{} asBias{}}\nolimits}
\newcommand{\argmin}{\mathop{\rm{} argmin{}}}
\newcommand{\tr}{\mathop{\rm{} tr{}}}
\newcommand{\B}{\mathbb B}

\newcommand{\R}{\mathbb R}

\newcommand{\EM} {{\mathbb I}}

\newcommand{\Tfrac}[2]{\textstyle\frac{#1}{#2}}

\newcommand{\nquad}{\!\!\!\!}
\newcommand{\mparagraph}[1]{\vspace{0.1ex}\noindent\textbf{#1}}%
\numberwithin{equation}{section}
\newcommand{\uQi}{\vspace*{-3ex}}
\newcommand{\uPi}{\vspace{-2.5ex}}
\newcommand{\uQii}{\vspace*{-.1ex}}
\newcommand{\uPii}{\vspace{-1ex}}
\newcommand{\uQiii}{\vspace*{-.1ex}}
\newcommand{\uPiii}{\vspace{-1ex}}
\newcommand{\uQiv}{\vspace*{-1.4ex}}
\newcommand{\uPiv}{\vspace{-4ex}}
\newcommand{\uQv}{\vspace*{-.5ex}}
\newcommand{\uPv}{\vspace{-1ex}}
\newcommand{\uP}{\vspace{-1ex}}
\newcommand{\uPTiii}{\vspace{-3ex}}
\newcommand{\Bullet}{{\centerdot}}

\renewcommand{\eqref}[1]{(\ref{#1})}
\RequirePackage{ifthen}

\begin{document}
%\doi{10.1080/0233188YYxxxxxxxx}
%\issn{1029-4910}
%\issnp{0233-1888}
%\jvol{00} \jnum{00} \jyear{2011} \jmonth{April}

%\markboth{Ruckdeschel \& Horbenko}{Statistics}

\articletype{Research article}

\title{{\itshape
\ifx\blinded\undefined
Optimally-Robust Estimators in\break Generalized Pareto Models
\thanks{This work was supported
by a DAAD scholarship for N.~Horbenko. %\newline
It is part of her PhD thesis, a preprint of it is Ruckdeschel and Horbenko~\citeyear{Ru:Ho:10}.
}}}

\author{Peter Ruckdeschel$^{\rm a}$ $^{\ast}$
\thanks{$^\ast$Peter Ruckdeschel. Email: peter.ruckdeschel@itwm.fraunhofer.de
\vspace{6pt}}
and Nataliya Horbenko$^{\rm a, b}$\\
\vspace{6pt}
$^{\rm a}${\em{Frauhofer ITWM, Kaiserslautern, Germany}}; \break
$^{\rm b}${\em{Kaiserslautern University, Germany}}
\\\vspace{6pt}
\received{v3.2 released April 2011}}
\maketitle
\fi

\begin{abstract}
We study robustness properties of several procedures for joint estimation of
shape and scale in a generalized Pareto model.
The estimators we primarily focus on, MBRE and OMSE, are one-step estimators
distinguished as optimally-robust in the shrinking neighborhood setting,
i.e.; they minimize the maximal bias, respectively, on
a specific such neighborhood, the maximal mean squared error.
For their initialization, we propose a particular
Location-Dispersion estimator, MedkMAD,
which matches the population median and kMAD
(an asymmetric variant of the median of absolute deviations)
 against the empirical counterparts.

These optimally-robust estimators are compared to maximum likelihood,
skipped maximum likelihood, Cram\'er-von-Mises minimum
distance, method of median, and Pickands estimators.
To quantify their deviation from  robust optimality, for each of these
suboptimal estimators, we determine the finite sample breakdown point,
the influence function, as well as the statistical accuracy
measured by asymptotic bias, variance, and mean squared error---all evaluated
uniformly on shrinking neighborhoods.
These asymptotic findings are complemented  by an extensive simulation study
to assess the finite sample behavior of the considered procedures.
Applicability of the procedures and their stability against outliers
is illustrated at the Danish fire insurance data set from {\sf R} package~{\tt evir}.
\bigskip

\begin{classcode}MSC 62F10, 62F35\end{classcode}\bigskip
\end{abstract}

%---------------------------------------------------------------------------------------------------------
\section{Introduction}
\label{intro}
%---------------------------------------------------------------------------------------------------------
This paper deals with optimally-robust parameter estimation in generalized
Pareto distributions (GPDs). These arise naturally in many situations
where one is interested in the behavior of extreme events as motivated
by the Pickands-Balkema-de\,Haan extreme value theorem (PBHT), cf.\
\citet{B:H:74}, \citet{Pick:75}.
The application we have in mind is calculation of the regulatory capital
required by \citet{ICCMCS:04} for a bank to cover operational risk, see
\citet{R:H:B:11}.
In this context, the tail behavior of the underlying distribution is crucial.
%Estimating these population quantiles by their empirical counterparts is drastically to outliers
%Nonparametric estimation of its high quantiles is drastically unstable against the outliers.
%
This is where extreme value theory enters, suggesting to estimate
these high quantiles parameterically using, e.g. GPDs, see \citet{N:C-D:E:06}.
%This per se is no remedy, however. Maximum Likelihood Estimators (MLEs), optimal
%in this parametric context, still attribute unbounded influence to some
%exposed observations.
%
%
Robust statistics in this context offers procedures bounding
the influence of single observations, so provides reliable inference
in the presence of moderate deviations from the distributional model assumptions,
respectively from the mechanisms underlying the PBHT.
%Admittedly, this comes at the price of some efficiency loss in the ideal model.

%---------------------------------------------------------------------------------------------------------
\mparagraph{Literature:}
%---------------------------------------------------------------------------------------------------------
%
Estimating the three-parameter GPD, i.e., with parameters for threshold,
scale, and shape, has been a challenging
problem for statisticians for long, with many proposed approaches.
In this context, estimation of the threshold is an important topic of its own
but not covered by the framework used in this paper. Here we rather limit
ourselves to joint estimation of scale and shape and
assume the threshold to be known.
In the meantime, for threshold estimation  we refer to
 \citet{B:V:T:96a,B:D:G:M:99}, while robustifications of this problem
 can be found in %\citet{V:B:H:04},
\citet{Du:98}, \citet{D:V-F:06}, and \citet{V:B:C:H:07}.

We also do not discuss non-parametric or semiparametric approaches
for modelling the tail events (absolute or relative excesses over the high threshold)
only specifying the tail index $\alpha$ through the number of exceedances over a
high threshold.
The most popular estimator in this family is the Hill estimator \citep{Hill:75};
for a survey on approaches of this kind, see \citet{T:P:01}.
With their semi/non-parametric nature, these methods can take into account the fact
that the GPD is only justified asymptotically by the PBHT and for finite samples is merely
a proxy for the exceedances distribution.
On the other hand, none of these estimators considers an unknown scale parameter
directly, but define it depending on the shape, so these estimators do
not fall into the framework studied in this paper.\smallskip

In parametric context, for estimation of scale and shape of a GPD,
the maximum likelihood estimator  (MLE) is highly popular among
practitioners, and has been studied in detail by \citet{Smith:87}.
This popularity is largely justified for the ideal model by the (asymptotic)
results on its efficiency, see \citet[Ch.~8]{vdW:98}, by which the
MLE achieves highest accuracy in quite a general setup.
\newline
The MLE looses this optimality however when passing over to only
slightly distorted distributions which calls for robust alternatives.
To study the instability of the MLE, \citet{C:M:A:U:09}
consider skipping some extremal data peaks, with the rationale to
 reduce the influence of extreme values. Grossly speaking, this
amounts to using a Skipped Maximum Likelihood Estimator (SMLE),
which enjoys some popularity among practitioners.
Close to it, but bias-corrected,
is the weighted likelihood method proposed in \citet{D:M:02}.
\citet{Du:98} %and \citet{D:F:98} for GEVD
studies optimally bias-robust estimators (OBRE)
as derived in \citep[2.4~Thm.~1]{Ha:Ro:86},  realized
as M-estimators. %\ACHTUNG{[TRUE?]}.
\newline%
Generalizing \citet{H:F:99} to the GPD case, \citet{P:W:01} propose
a method of medians estimator, which is based on solving the
implicit equations matching the population medians
of the scores function to the data coordinatewise.
\newline%
Pickands estimator (PE) \citep{Pick:75} matches certain
empirical quantiles against the model ones and strikes out for its
closed form representation. This idea has been
generalized to the Elementary Percentile Method (EPM) by \citet{C:H:97}.
%
% \newline%
% \citet{B:S:00} use a different parametrization of the GPD,
% i.e. instead of observations $X_i \iid{\rm GPD}(\beta,\xi)$ in our notation,
% consider observations $Y_i=X_i+\beta/\xi$
% and parametrize their model by $\sigma=\xi$ and $\alpha= \log(\beta/\xi)$. In
% their setting,  ${\cal L}(\log(Y_i))={\cal L}(\alpha+\sigma E)$, $E\sim{\rm Exp}(1)$,
% transforming the problem to a location-scale problem for the exponential
% distribution. In our setting though, their procedures do not apply directly,
% as the transformation of the observation involves unknown
% parameters and %, as $\beta/\xi$ is unknown;
% in addition the location-scale model with ${\rm Exp}(1)$ as center is
% not $L_2$-differentiable, so not covered by standard theory. %, as it is not $L_2$-differentiable, which heuristically may be seen from
% the fact that observations close its to lower end-point $\alpha$ are
% extremely informative for its estimation.
%\newline%
%%%%%%%%%%%Nataliya: more estimators from literature added
\newline
Another line of research may be grouped into moments-based
estimators, matching empirical (weighted, trimmed) moments
of original or transformed observations against their model counterparts.
For the first and second moments of the original observations this
gives the Method of Moments (MOM), %estimator studied in \citet{H:W:87} \ACHTUNG{TRUE?};
for the probability-transform scaled observations this leads to
Probability Weighted Moments (PWM), see \citet{H:W:87};
a hybrid method of these two is studied in  \citet{D:T:98};
with the likelihood scale, this gives Likelihood Moment Method (LME)
as in \citet{Zh:07}.
\citet{B:K:09} cover trimmed moments.
Clearly, except for the last one, all these methods are restricted to cases
where the respective population moments are finite,
which may preclude some of them for certain applications:
for the operational risk data even first moments may not
exist \citep{N:C-D:E:06} so ordinary MOM
estimators cannot be used in these cases.
\newline%
Examples of minimum distance type estimators like the
Minimum Density Power Divergence Estimator (MDPDE)
or the Maximum Goodness-of-Fit Estimator (MGF) can
be found in  \citet{J:S:04} and \citet{Lu:06}, respectively.
%In this paper we study a minimum distance estimator
%based on Cram\'er-von-Mises distance.
%
%
%We do not study these estimators in details further.
%\smallskip

%---------------------------------------------------------------------------------------------------------
\mparagraph{Considered estimators: }
%---------------------------------------------------------------------------------------------------------
%This paper covers %\vspace{-1.3ex}
%\begin{itemize}
%  \item the Maximum Likelihood Estimator (MLE)
%  \item the Skipped Maximum Likelihood Estimator (SMLE)
%  \item the Cram\'er-von-Mises Minimum Distance estimator (MDE)
%  \item Pickands Estimator (PE)
%  \item the Method-of-Median estimator (MMed)
%  \item an estimator based on median and kMAD (MedkMAD)
%  \item the most bias-robust estimator minimizing the maximal bias (MBRE)
%  \item the estimator minimizing the maximal MSE,
%when the radius of contamination is known (OMSE) / not known (RMXE) %\vspace{-1.8ex}
%\end{itemize}
%For actual definitions see section~\ref{Estimators}.
%This choice is motivated as follows: MLE, MBRE, OMSE, RMXE are optimal in
Except for \citet{Du:98}, non of the mentioned
 robustifications heads for robust optimality. This is the topic
 of this paper. In the GPD setup, we study estimators distinguished as optimal, i.e.,
 the maximum likelihood estimator (MLE),
the most bias-robust estimator minimizing the maximal bias (MBRE),
and the estimator minimizing the maximal MSE on gross error neighborhoods
about the GPD model, when the radius of contamination is known (OMSE)
and not known (RMXE). %\\
These estimators need globally-robust initialization estimators; for this purpose
we consider Pickands estimator (PE), the method-of-median estimator (MMed)
and a particular Location-Dispersion (LD) estimator, MedkMAD. %\\
From our application of these estimators to operational risk,
we take the skipped maximum likelihood estimator (SMLE)
and  the Cram\'er-von-Mises Minimum Distance estimator (MDE) as
competitors.

%---------------------------------------------------------------------------------------------------------
\mparagraph{Contribution of this article: }
%---------------------------------------------------------------------------------------------------------
Our contribution is a translation of asymptotic optimality from \citet{Ried:94}
to the GPD context and derivation of the optimally-robust estimators MBRE, OMSE, and RMXE in this
context together with their equivariance properties in Proposition~\ref{scalinvprop}.
This also comprises an actual implementation to determine the respective
influence functions in {\sf R}, including a considerable speed-up
by interpolation with Algorithm~\ref{algor}. Moreover, for initialization
of MLE, MBRE, OMSE, RMXE, we propose a computationally-efficient starting
estimator with a high breakdown---the MedkMAD estimator,
which improves known initialization-free estimators considerably. For its
distinction from alternatives, common finite sample breakdown point
notions to assess global robustness have to be replaced by the
concept of expected finite sample breakdown point
introduced in \citet{Ru:Ho:10b}.
While the optimality results of \citet{Ried:94} do not quantify
suboptimality of competitor estimators, our synopsis in
Section~\ref{Synopsis} provides a detailed discussion of this issue.
To this end, in Appendix~\ref{EstimatorsApp}, in Propositions~\ref{MLEprop}%,
%\ref{SMLEprop}, \ref{MDEprop}, \ref{PEprop}, \ref{MMedprop}, and
--\ref{MedkMADprop}, we provide a variety
of largely unpublished results on influence functions, asymptotic
(co)variances, (maximal) biases, and breakdown points of the
considered estimators.
The optimality theory we use is confined to an asymptotic framework
for sample size tending to infinity; the simulation results of
Section~\ref{Simulation Study} however close this gap by establishing
finite sample optimality down to sample size $40$.
%---among the considered competitors where
%each estimator has to prove its usefulness in its individual worst contamination situation.
%Finally, in Appendix~\ref{EstimatorsApp}, we provide a variety
%of results on influence functions, asymptotic (co)variances, (maximal) biases,
%and breakdown points of the considered estimators.
%, most of them unpublished before.
%

%---------------------------------------------------------------------------------------------------------
\mparagraph{Structure of the paper:}
%---------------------------------------------------------------------------------------------------------
In Section~\ref{Model Setting} we define the ideal model and summarize its smoothness and invariance properties,
and then extend this ideal setting  defining contamination neighborhoods.
Section~\ref{Robustness} provides basic global and local robustness concepts
and recalls the influence functions of optimally robust estimators;
it also introduces several efficiency concepts. Section~\ref{Estimators} introduces the
considered estimators, discusses some computational and numerical aspects and
 in a synopsis summarizes the respective robustness properties.
A simulation study in Section~\ref{Simulation Study} checks for the validity of the
asymptotic concepts at finite sample sizes. To illustrate the stability of the considered
estimators at a real data set, in Section~\ref{Appsec}, we evaluate the estimators at the
Danish fire insurance data set of {\sf R} package {\tt evir} \citep{M:S:07} and at a
modified version of it, containing $1.5\%$ outliers.
Our conclusions are presented in Section~\ref{concl}.
Appendix~\ref{EstimatorsApp} provides our calculations behind our
results in the synopsis section. Proofs are provided in Appendix~\ref{proofsec}.

%---------------------------------------------------------------------------------------------------------
\section{Model Setting}\label{Model Setting}
%---------------------------------------------------------------------------------------------------------
\subsection{Generalized Pareto Distribution}
%---------------------------------------------------------------------------------------------------------
The three-parameter generalized Pareto distribution (GPD) has c.d.f.\ and density
\begin{eqnarray} \label{GPDmodel}
&&F_{\theta}(x) = 1- \left( 1 + \xi \frac{x - \mu}{\beta} \right)^{-\frac{1}{\xi}}, \quad
f_{\theta}(x) = \frac{1}{\beta}
\left( 1 + \xi \frac{x - \mu}{\beta} \right)^{-\frac{1}{\xi} - 1}
\end{eqnarray}
where $x \geq \mu$ for $\xi \geq 0$, and $\mu < x \leq \mu - \frac{\beta}{\xi}$ if $\xi < 0$.
It is parametrized by $\vartheta=(\xi,\beta,\mu)^\tau$, for location $\mu$, scale $\beta>0$ and shape $\xi$.
Special cases of GPDs are the uniform ($\xi = -1$), the exponential ($\xi = 0$, $\mu=0$), and
Pareto ($\xi > 0$, $\beta=1$) distributions.

We limit ourselves to
the case of %shape $\xi>0$ and
known location $\mu=0$ here; for shape values of $\xi>0$,
GPD is a good candidate for modeling distributional tails exceeding threshold $\mu$
as motivated by the PBHT, but for simplicity we do not make this restriction in
this paper; with this restriction, corresponding log-transformations as discussed
later for scale $\beta$ would also be helpful for shape $\xi$.
For all graphics and both numerical evaluations and simulations,
we use the reference parameter values $\beta=1$ and $\xi=0.7$.
For known $\mu$, the model is smooth in $\theta=(\xi,\beta)$:

\begin{proposition} \label{smoothmodel}
For given $\mu$ and at any $\xi\in\R$, $\beta > 0$, the GPD model from \eqref{GPDmodel}
is $L_2$-differentiable w.r.t.\ $(\beta, \xi)$, with
$L_2$-derivative (or scores)
\begin{equation} \label{LBdef}
\Lambda_\theta(z) = \left( \Tfrac{1}{\xi^2}\log(1+\xi z) -
\Tfrac{\xi+1}{\xi}\Tfrac{z}{1+\xi z};-\Tfrac{1}{\beta}+
\Tfrac{\xi+1}{\beta}\Tfrac{z}{1+\xi z}\right)^\tau, \quad
z = \Tfrac{x-\mu}{\beta}
\end{equation}
and finite Fisher information
${\mathcal{I}}_\theta$
\begin{equation}\label{FIdef}
\mathcal{I}_\theta = \frac{1}{(2\xi+1)(\xi+1)}
\left(\begin{array}{cc}
 2, & \beta^{-1} \\
\beta^{-1}, & \beta^{-2} (\xi+1)
\end{array}\right) \succ 0
\end{equation}
\end{proposition}
As ${\cal I}_\theta$ is positive definite for $\xi\in\R$, $\beta>0$,
the model is (locally) identifiable.

\mparagraph{In-/Equivaraince}
The model for given $\mu$ is \textit{scale invariant} in the sense that  for $X$
a random variable (r.v.) with law ${\cal L}(X)=F_{(\xi,b,\mu)}$, for $\beta>0$ also
${\cal L}(\beta X)=F_{(\xi,b\beta,\mu)}$ is in the model.
Using matrix $d_\beta=\mathop{\rm diag}(1,\beta)$, correspondingly, an estimator
$S$ for $\theta=(\xi,\beta)$ is called \textit{(scale)-equivariant} if
\begin{equation}\label{scaleeq}
S(\beta x_1,\ldots, \beta x_n)= d_\beta S(x_1,\ldots, x_n)
\end{equation}
However, no such in-/equivariance is  evident for the shape component.

Later on, it turns out useful to transform the scale parameter to logarithmic scale,
because of breakdown of scale estimates, see Lemma~\ref{Lemma_exp_scale} below,
i.e.; to estimate $\tilde \beta=\log\beta, \ \beta = e^{\tilde \beta}$ and then, afterwards to back-transform the estimate
to original scale by the exponential. For the transformed model, we write
\begin{equation} \label{logtrafo1}
\tilde \beta=\log\beta,\qquad \tilde \theta=(\xi,\tilde \beta),\qquad \tilde \Lambda_{\tilde \theta}(z)=\frac{\partial}{\partial\tilde\theta} \log f_\theta(z),\qquad
\tilde {\cal I}_{ \tilde  \theta}= \Ew_{\tilde \theta} \tilde \Lambda_{\tilde \theta} \tilde \Lambda_{\tilde \theta}^\tau
\end{equation}

On log-scale, scale equivariance
\eqref{scaleeq} translates into a shift equivariance:
an estimator
$\tilde S$ for $\tilde \theta=(\xi,\tilde \beta)$ is called \textit{(shift)-equivariant} if
\begin{equation}\label{shifteq}
S(\beta x_1,\ldots, \beta x_n)= S(e^{\tilde \beta} x_1,\ldots, e^{\tilde \beta} x_n)= S(x_1,\ldots, x_n) + (0,\tilde \beta)^\tau
\end{equation}

\begin{lemma}\label{InvProp}
For the scores these invariances are reflected by the relations
\begin{equation}
\Lambda_{\theta}(x)=d_\beta^{-1} \Lambda_{\theta_1}(\Tfrac{x}{\beta}),\quad\;\;
\mathcal{I}_{\theta}=d_\beta^{-1} \mathcal{I}_{\theta_1}d_\beta^{-1},\qquad\quad
\tilde\Lambda_{\tilde\theta}(x) =\tilde\Lambda_{\tilde\theta_0}(\Tfrac{x}{\beta}),\quad\;\;
\tilde {\cal I}_{\tilde \theta} = \tilde {\cal I}_{\tilde\theta_0} \label{InvProp1}
\end{equation}
where
\begin{equation} \label{th1th0def}
\theta_1=(\xi,1)\qquad \mbox{respectively}\qquad \tilde \theta_0=(\xi,0)
\end{equation}
and
\begin{equation} \label{Lbtransf}
\tilde\Lambda_{\tilde\theta}(x)=d_\beta\Lambda_{\theta}(x)
\end{equation}
\end{lemma}
%

%To preserve this (partial) invariance when determining the ``length''
%of a parameter, robust statistics uses special norms for
%the parameter space; as a simple scale invariant norm, we
% use the weighted norm
%%
%\begin{equation}
%n_\beta(x,y)=\|d_\beta^{-1}(x,y)\|=\sqrt{x^2+y^2/\beta^2} \label{nbetdef}
%\end{equation}
%%

% \begin{Rem}
% For the shape parameter there is no obvious such invariance,
% except for the quantile transformation, of course, i.e.
% %
% \begin{equation} \label{equivar1}
% g(\theta,\theta';x)=F_{\theta'}^{-1}\circ F_\theta (x) =
% \big[(1+{\xi}x/{\beta})^{\xi'/\xi}-1\big]{\beta'}/{\xi'}
% \end{equation}
% %
% transforming an $F_\theta$-distributed observation
% $X$ into an $F_{\theta'}$-distributed one. The only values of $x$
% invariant under arbitrary $g(\theta,\theta';\,\cdot\,)$ are
% $\{0,\infty\}$, as in the pure scale case.
% However, with this group, we do not see any form of reasonable
% equivariance.
% \end{Rem}

%---------------------------------------------------------------------------------------------------------
\subsection{Deviations from the Ideal Model: Gross Error Model}
%---------------------------------------------------------------------------------------------------------
Instead of working only with ideal distributions, robust statistics
considers suitable distributional neighborhoods about this ideal model.
In this paper, we limit ourselves to the \textit{Gross Error Model},
i.e. our neighborhoods are the sets of all real distributions $F^{\ssr re}$
 representable as
\begin{equation}\label{GEM}
F^{\ssr re}=(1-\ve) F^{\ssr id} + \ve F^{\ssr di}
\end{equation}
for some given size or radius $\ve>0$,
where $F^{\ssr id}$ is the underlying ideal distribution and $F^{\ssr di}$ some
arbitrary, unknown, and uncontrollable contaminating/distorting distribution which may
vary from observation to observation.
For fixed $\ve>0$, bias and variance of robust estimators usually
scale at different rates ($\LO(\ve)$, $\LO(1/n)$, respectively).
Hence to balance bias and variance scales, in the shrinking neighborhood approach, see
\citet{HC70}, \citet{Ri78,Ried:94}, and \citet{Bic:81}, one
lets the radius of these neighborhoods shrink with growing sample
size $n$, i.e.
\begin{equation} \label{shrinkRate}
\ve = r_n= {r}/{\sqrt{n}}
\end{equation}
%(and contamination $F^{\ssr di}$ may vary from observation
%to observation and in $n$  as well).

In reality one rarely knows $\ve$ or $r$, but for situations
where this radius is not exactly known, in \citet{R:K:R:08}
we provide a criterion to choose a radius then; this is detailed
in Section~\ref{EffSec}.
Our reference radius for our evaluations and simulations is $r = 0.5$.
%
%\begin{Rem}\label{RemMinMax}
%  \small \rm
%$r=0.5$
%The corresponding maximin efficiency is $0.683$ compared to
%a minimal $0.678$ efficiency of the procedure tuned for radius
%$0.5$.
%, i.e. using the
%respective radius minimax procedure,
%the performance of this procedure would never be worse than
%$1.464$ times the maximal $\asMSE$ (see below) of the optimal procedure knowing the radius.
%The minimal efficiency of the OMSE to radius $r=0.5$ is in
%fact only $0.678$ (achieved when used for unknown radius $r=0$).
%\end{Rem}
%---------------------------------------------------------------------------------------------------------
\section{Robust Statistics}\label{Robustness}
%---------------------------------------------------------------------------------------------------------

To assess robustness of the considered estimator against these deviations,
we study local properties measuring the infinitesimal influence of a single observation
as the \textit{influence function} (IF) and global ones like the \textit{breakdown point}
measuring the effect of massive deviations.

%..............................
\subsection{Local Robustness: Influence Function and ALEs}
%..............................
%
%Defining an estimator as a functional $T$ evaluated at the empirical
%distribution, the IF of $T$ is the functional
%derivative of the estimator with respect to the distribution.
%Historically, in  is defined as the
%G\^ateaux derivative %in the direction of a Dirac measure
%(provided the limit exists):
For $\delta_x$ the Dirac measure at $x$ and
$F_\ve =(1-\ve) F + \ve \delta_x$, \citet{Ha:68}
defines the influence function of a statistical functional $T$
at distribution $F$ and in $x$ as
\begin{equation}
\IF(x;T,F) = \lim_{\ve \rightarrow 0}\frac{T(F_\ve)-T(F)}{\ve}
\end{equation}
provided the limit exists.
\citet[(introduction)]{K:R:R:09} summarize some pitfalls of this
definition, which in our context however can
be avoided:  by the Delta method, this amounts
to the question of Hadamard differentiability
of the likelihood (MLE, SMLE), of quantiles (PE, MMed,
MedkMAD), and of the c.d.f.\ (MDE). Indeed, results from \citet{Fernholz},
\citet[Ch.~1,6]{Ried:94} establish that all our estimators
 are ALEs in the following sense.

%------------------------------------
\mparagraph{ALEs}
%------------------------------------
\textit{Asymptotically linear} estimators
or \textit{ALE}s in our GPD model are estimators $S_n$ for parameter $\theta$,
 having the expansion in the observations $X_i$ as
\begin{equation} \label{lunifn}
S_n=\theta + \frac{1}{n} \sum_{i=1}^n \psi_\theta(X_i) + R_n,
\qquad \sqrt{n}\,|R_n| \stackrel{n\to\infty}{\longrightarrow} 0
\quad\mbox{$P_{\theta}^n$-stoch.}
\end{equation}
for $\psi_\theta\in L_2^2(P_\theta)$ the IF of $S_n$ for which we require
\begin{equation} \label{ICcond}
\Ew_\theta \psi_\theta =0,\qquad
\Ew_\theta \psi_\theta \Lambda_\theta^\tau={\mathbb I}_{2}
\end{equation}
(with ${\mathbb I}_{2}$ the $2$-dim.\ unit matrix and $L_2^2(P_\theta)$ the
set of all $2$-dim.\ r.v.'s $X$ s.t.\ $\int |X|^2\, dP_\theta<\infty$).\newline
Note that for \eqref{ICcond} we need $L_2$-differentiability 
as shown in Proposition~\ref{smoothmodel}.\newline
Using \eqref{Lbtransf} one easily sees that if $\psi_\theta$ is an IF in
the model with original scale,
\begin{equation}
\eta_{\tilde\theta}(x):=d_\beta^{-1}\psi_\theta(x) \label{11corresp}
\end{equation} is
an IF in the log scale model, so there is a one-to-one correspondence between the IFs in
these models.

In the sequel we fix the true parameter value $\theta$ and suppress the respective
subscript where unambiguous.
The class of all $\psi \in L^2_2(P)$  satisfying \eqref{ICcond} is denoted by $\Psi_2$.
%Equation~\eqref{ICcond} may be motivated either by \citet[Lem.~4.2.18]{Ried:94}
%or \citet[Lem.~1.3]{Ru:Ho:10}.
%An estimator with \eqref{lunifn} is called
%\textit{asymptotically linear} or \textit{ALE}.
%We note that all estimators considered in this paper are ALEs.
%
In the class
of ALEs asymptotic variance and the maximal
asymptotic bias may be expressed in terms of the respective IF only,
as recalled in the following proposition.

\begin{proposition}\label{ALEprop}
Let ${\cal U}_n$  be a sequence of shrinking
neighborhoods in the gross error model~\eqref{GEM}, \eqref{shrinkRate} with starting radius $r$.
Consider an ALE $S_n$  with IF $\psi$.
The ($n$-standardized) asymptotic (co)variance matrix of $S_n$ on ${\cal U}_n$  is given by
\begin{equation}
\asVar(S_n) = \int \psi\psi^{\tau} \,dF
\end{equation}
The $\sqrt{n}$-standardized, maximal asymptotic bias  $\asBias(S_n)$ on ${\cal U}_n$ is  $ r\,\cdot\, {\rm GES}(\psi)$
 where
\begin{equation} \label{GESdef}
{\rm GES}(\psi):= \sup\nolimits_{x}|\psi(x)| %\vspace{-1ex}
\end{equation}
is the \textit{gross error sensitivity} and $|\,\cdot\,|$ is the Euclidean norm.
The (maximal, $n$-standardized) asymptotic mean squared error (MSE) $\asMSE(S_n)$ on
${\cal U}_n$  is given by
\begin{equation} \label{asMSEdef}
\asMSE(S_n) = r^2 \,{\rm GES}^2 + \tr(\asVar(S_n))
\end{equation}
\end{proposition}
For a proof of this proposition we refer to \citet[Rem.~4.2.17(b), Lem.~5.3.3]{Ried:94};
for the notion ``gross error sensitivity'' see \citet[Ch. 2.1c]{Ha:Ro:86}.

%------------------------------------
\mparagraph{Optimally-robust ALEs}
%------------------------------------
By Proposition~\ref{ALEprop} we may delegate optimizing robustness to the class of IFs;
the optimally-robust IFs are determined in the following proposition due
to \cite[Thm.'s 5.5.7 and 5.5.1]{Ried:94}.

\begin{proposition} \label{OptRobProp}
In our GPD model enlarged by \eqref{GEM}, \eqref{shrinkRate}, 
the unique ALE minimizing $\asBias$, denoted by \emph{MBRE}, is given by its IF $\bar \psi$ where
$\bar\psi$ is necessarily of form
\begin{equation} \label{barpsidef}
\bar \psi = b{Y}/{|Y|}, \quad Y = A \Lambda - a, \quad b =
\max_{a,A}\{\tr(A)/{\Ew|Y|}\}\;,
\end{equation}
and the  unique ALE minimizing $\asMSE$ on a (shrinking) neighborhood of radius $r$, denoted by \emph{OMSE} is given by its IF $\hat \psi$
where $\hat\psi$ is necessarily of form
\begin{equation} \label{hatpsidef}
\hat{\psi} = Y \min \left\{1,{b}/{|Y|}\right\}, \quad Y = A \Lambda - a,
\quad r^2 b = \Ew(|Y|-b)_+ \;.
\end{equation}
In both cases $A \in\R^{2\times 2}$, $a\in\R^2$, $b>0$ are Lagrange multipliers
ensuring that  $\psi \in\Psi_2$.
\end{proposition}

\textit{Invariance}
Lemma~\ref{InvProp} entails an invariance of the optimally-robust IFs,
which allows a reduction to reference scale $\theta_1$ respectively $\tilde \theta_0$
from \eqref{th1th0def} and alleviates computation considerably---provided
in the original ($\beta$-)scale model, we replace Euclidean norm $n_1$ by
\begin{equation}
n_\beta(x):=|d_\beta^{-1} x|=\sqrt{x_1^2+x_2^2/\beta^2\,}
\end{equation}
In particular, by correspondence~\eqref{11corresp} the optimal solutions
in original scale and in log-scale coincide.
\begin{proposition}\label{scalinvprop}
\begin{itemize}
\item[(a)] Replacing Euclidean norm  by $n_\beta$ in Proposition~\ref{OptRobProp}, the optimal IFs are as
in \eqref{barpsidef} and \eqref{hatpsidef}, where one has to replace
expression $\tr(A)$ by $\tr (d_\beta^{-2}A)$ in \eqref{barpsidef}.
\item[(b)] In the original scale model, with norm $n_\beta$, for $\psi = \hat \psi$ or $\psi=\bar\psi$,
\begin{equation} \label{bet=1lsg}
\psi_{\theta}(x)=d_\beta \psi_{\theta_1}(x/\beta)
\end{equation}
and the Lagrange multipliers translate according to
\begin{equation}
A_{\theta}
=d_\beta A_{\theta_1}d_\beta, \quad\;\;
a_{\theta}
=d_\beta a_{\theta_1},\quad\;\;
b_{\theta}
=b_{\theta_1} \label{Aeqvar} %,\qquad %\\
%&&
% \label{Yeqvar}
\end{equation}
\item[(c)] In the log-scale model with the Euclidean norm, the Lagrange multipliers remain invariant under parameter
changes and writing $\eta$ for the optimal IFs,
\begin{equation}
\eta_{\tilde \theta}(x)=\eta_{\tilde \theta_0}(x/\beta)
\end{equation}
\item[(d)] The optimally-robust IFs with their Lagrange multipliers $\tilde A$, $\tilde a$, and $\tilde b$ in the log-scale model from (c) are related to the ones in
the original scale from (b) by
\begin{equation} \label{etatranspsi}
\eta_{\tilde \theta}(x)=d_{\beta}^{-1}\psi_{\theta}(x),\qquad \tilde A = d_\beta^{-1}A_\theta d_\beta^{-1},\;\;\;\tilde a=d_\beta^{-1} a_\theta,\;\;\;\tilde b= b_\theta
\end{equation}
\end{itemize}
\end{proposition}

In a subsequent construction step, one has to find an ALE achieving the optimal IF.
For this purpose, we use the one-step construction, i.e.; to a suitable
starting estimator $\theta_n^{(0)}=\theta_n^{(0)}(X_1,\ldots,X_n)$
and IF $\psi_\theta$, we define
\begin{equation}\label{onestep}
S_n = \theta_n^{(0)} + \frac{1}{n}\sum_{i=1}^n \psi_{\theta_n^{(0)}}(X_i)
\end{equation}
For exact conditions on $\theta_n^{(0)}$ see \citet[Ch.~6]{Ried:94} or \citet[Sec.~2.3]{KohlDiss}.
Suitable starting estimators allow to interchange supremum and integration,
and $\asMSE$ also is the standardized asymptotic maximal MSE.

%..............................
\subsection{Global Robustness: Breakdown Point}
%..............................
The breakdown point in the gross error model \eqref{GEM} gives the largest
radius $\ve$ at which the estimator still produces reliable results. We take the
definitions from \citet[2.2~Definitions~1,2]{Ha:Ro:86}.
\noindent
The \emph{asymptotic breakdown point (ABP)}  $\ve^\ast$ of the sequence of
estimators $T_n$ for parameter $\theta\in\Theta$ at probability $F$ is given by
%\begin{eqnarray}
\begin{equation}
\!\ve^\ast\!:=\!\sup\Big\{\ve \!\in\! (0,1]\,\Big| \exists \;\mbox{\small compact}\,K_\ve \!\subset\! \Theta\colon
%\mbox{there is a compact set
%$K_\ve \subset \Theta$ s.t.}\nonumber\\
%&&
\pi(F,G) \!<\! \ve\,\Rightarrow\, G(\{T_n \!\in\! K_\ve\})
\stackrel{n\to\infty}{\rightarrow} 1\Big\},\!\!
\end{equation}
%\end{eqnarray}
where $\pi$ is Prokhorov distance.
\noindent
The \emph{finite sample breakdown point (FSBP)} %\label{def:fin_breakdown_point}
$\ve_n^\ast$ of the estimator $T_n$ at the sample $(x_1,...,x_n)$ is given by
\begin{equation} \label{FSBPDef}
\ve_n^*(T_n; x_1,...,x_n) := \frac{1}{n} \max \Big\{ m;
\max_{i_1,...,i_m} \sup_{y_1,...,y_m}|T_n(z_1,...,z_n)| < \infty \Big\},
\end{equation}
where the sample $(z_1,...,z_n)$ is obtained by replacing the %$m$
data points $x_{i_1},...,x_{i_m}$ by arbitrary values $y_1,...,y_m$.
%
%The  ABP was introduced in \citet{Ha:68}, and
%the FSBP in \citet{Do:Hu:83}, but note that $\ve_n^\ast$ from  \eqref{FSBPDef} is by $1/n$
%smaller than the Donoho-Huber one.
Definition~\eqref{FSBPDef} however does not cover
implosion breakdown of scale parameter. Passage to the log-scale as in \eqref{logtrafo1}
provides an easy remedy though, compare \citet{He:05}, i.e.;
%\begin{Def}
\begin{equation} \label{BPdef2}
\ve_n^\ast(T_n; x_1,...,x_n) := \frac{1}{n} \max \Big\{ m;
\max_{i_1,...,i_m} \sup_{y_1,...,y_m}|\log(T_n(z_1,...,z_n))| < \infty \Big\}.
\end{equation}

\mparagraph{Expected finite sample breakdown point }
For deciding upon which procedure to take \textit{before} having made
observations, in particular for ranking procedures in a simulation study,
the FSBP from \eqref{FSBPDef} has some drawbacks: for some of the considered
estimators, the dependence on possibly highly improbable configurations of
the sample entails that not even a non-trivial lower bound for the FSBP
exists. To get rid of this dependence to some extent at least, but still
preserving the finite sample aspect, we use the supplementary notion
of \textit{expected} FSBP (EFSBP) proposed and discussed in detail
in \citet{Ru:Ho:10b}, i.e.;
\begin{equation}
\bar \ve_n^*(T_n) := \Ew \ve_n^*(T_n; X_1,...,X_n)
\end{equation}
where expectation is evaluated in the ideal model.
We also consider the limit %
$\bar \ve^*(T) := \lim_{n\to \infty} \bar \ve_n^*(T_n)$
and also call it EFSBP where unambiguous.

%\begin{Rem}
\mparagraph{Inheritance of the breakdown point }
If the only possible parameter values where breakdown occurs are at infinity,
it is evident from equation~\eqref{onestep} that for bounded IF, a one-step estimator
inherits the breakdown properties of the starting value $\theta_n^{(0)}$.
This is not true for scale parameter $\beta$.
If scale component $\beta^{(0)}_n>0$ of the starting estimate
$\theta^{(0)}_n$ is small, it can easily happen that the scale
component of the one-step construction fails to be positive, entailing
an implosion breakdown.
%\end{Rem}
%
Lemma~\ref{Lemma_exp_scale} below shows that we avoid this, if,
in the one-step construction, we pass to log-scale as in
\eqref{logtrafo1} (and afterwards back-transform); %, we
%\eqref{expscale}
%avoids this;
in the lemma,
 we write $\psi_2(x;\theta)$ for the scale component
 of IF $\psi_\theta(x)$ (in the untransformed model) evaluated at observation $x$ and parameter $\theta$.

\begin{lemma}\label{Lemma_exp_scale}
Consider construction~\eqref{onestep} with
starting estimator $S_n^{(0)}=(\beta_n^{(0)},\xi_n^{(0)})^\tau$.
If scale part $\beta_n^{(0)}>0$ and if
  $\sup_x |\psi_2(x;S_n^{(0)})|=b<\infty$, for scale part
$\beta_n$ of one-step estimator $S_n$ back-transformed from log-scale, we obtain
\begin{equation} \label{expscale}
\beta_n=\beta_n^{(0)} \exp\Big( \frac{1}{n\beta_n^{(0)}} \sum_i \psi_2 (X_i;S_n^{(0)})\Big)>0
\end{equation}
and the breakdown point of $\beta_n$ is equal to the one of $\beta_n^{(0)}$.
\end{lemma}
\vspace{-1.ex}

%..............................
\subsection{Efficiency}\label{EffSec}
%..............................
To judge the accuracy of
%a robust estimator $S_n$
an ALE $S = S_n$ it is natural to compare it
to the best achievable accuracy, giving  its (asymptotic relative) efficiency
{\rm eff.}\underline{\rm id} (in the \underline{id}eal model) defined as
\begin{equation}
{\rm eff.id}(S)=\frac{\tr(\asVar({\rm MLE})))}{\tr(\asVar(S))}=\frac{\tr({\cal I}^{-1})}{\tr(\asVar(S))}
\end{equation}
In terms of sample size $n$, (asymptotically) the optimal
estimator, i.e., the MLE in our case,  needs  $n \cdot (1-{\rm eff.id}(S))$
less observations to achieve the same accuracy as $S$.

Preserving this sample size interpretation,
 we extend this efficiency notion to situations under contamination of known radius
 $r$ (or \underline{re}alistic conditions)
{\rm eff.}\underline{\rm re}, defined again as a ratio w.r.t.\ the optimal procedure, i.e.,
\begin{equation}
{\rm eff.re}(S)={\rm eff.re}(S;r)=\frac{\asMSE({\rm OMSE}_r)}{\asMSE(S)}
%=\frac{\asMSE({\rm OMSE})}{\asMSE(S)}
\end{equation}
Finally, in \citet{R:K:R:08}, for the situation where \underline{r}adius $r$ is (at least partially)
\underline{u}nknown, we also compute the least favorable efficiency {\rm eff.}\underline{\rm ru}
\begin{equation}
{\rm eff.ru}(S):=\min_r{\rm eff.re}(S;r)
\end{equation}
where $r$ ranges in a set of possible radius values (here $r\in [0,\infty)$).
The radius $r_0$  maximizing ${\rm eff.ru}$ is called  \textit{least favorable radius}.
In our reference setting, i.e., for $\xi=0.7$ and $\beta=1$, we obtain
$r_0=0.486$ which is in fact very close to our chosen reference radius of $0.5$.

The procedure we recommend in this setting is the OMSE to $r=r_0$,
 called \textit{\underline{r}adius \underline{m}a\underline{x}imin \underline{e}stimator}
(RMXE); it achieves maximin efficiency {\rm eff.re}.

\begin{Remark} \label{Rem35}
It is common in robust statistics to use high breakdown point
estimators improved in a \textit{reweighting step} and tuned to achieve a
high efficiency ${\rm eff.id}$, usually to 95\%.
This practice to determine the degree of robustness is called \textit{Anscombe criterion}
and has its flaws, as the ``insurance premium'' paid
in terms of the  $5\%$ efficiency loss does not reflect the protection
``bought'', as this protection will vary \mbox{model-}, and in our
non-invariant case even $\theta$-wise. Instead, we recommend criteria
${\rm eff.re}$ and ${\rm eff.ru}$ to determine the degree of robustness.

Illustrating this point, in the GPD model at $\xi=0.7$, tuning the ${\rm OBRE}$
for ${\rm eff.id}=95\%$, where we indicate this tuning by a respective index for ${\rm OBRE}$, we
obtain
\begin{small}
\begin{eqnarray*}
&&\quad{\rm eff.id}({\rm OBRE}_{95\%})=95\%, \quad \mbox{but}\quad {\rm eff.ru}({\rm OBRE}_{95\%})=14\%,\\
&\mbox{while}&\quad {\rm eff.id}({\rm OMSE}_{r=0.5}) = {\rm eff.ru}({\rm OMSE}_{r=0.5})=67.8\%\\
&\mbox{and}&\quad  {\rm eff.id}({\rm RMXE})={\rm eff.ru}({\rm RMXE})=68.3\%,
\end{eqnarray*}
\end{small}
These $14\%$ indicate an unduely high vulnerability of ${\rm OBRE}_{95\%}$ w.r.t.\ bias.
For plots of the curve $r\mapsto {\rm eff.re}(S;r)$ we refer to \citet[p.26]{R:K:R:08}
(up to using reciprocal values for relative efficiencies);
as shown there, the curve is bowl-shaped, decreasing for $r\to 0,\infty$;  ${\rm OBRE}_{95\%}$ takes its minimum for $r=\infty$,
while for RMXE both local minima, i.e., at $r=0$ and $r=\infty$ are equal.
\end{Remark}
%-------------------------------------------------------------------------------------------------------
\section{Estimators}\label{Estimators}
%----------------------------------------------------------------------------------
In this section we gather the definitions of the
estimators considered in this paper; all of them are scale-invariant
(respectively shift-invariant passing to the log-scale); 
their robustness properties are
detailed in Appendix~\ref{EstimatorsApp} and summarized in Subsection~\ref{Synopsis}.
 %
%----------------------------------------------------------------------------------
\subsection{Optimal Estimators}
%We start with MLE-type estimators.

%----------------------------------------------------------------------------------
\mparagraph{MLE}
%----------------------------------------------------------------------------------
The maximum likelihood estimator is the maximizer (in $\theta$) of the
(product-log-) likelihood $l_n(\theta;X_1,\ldots,X_n)$ of our model
\begin{equation}
l_n(\theta;X_1,\ldots,X_n)=\sum_{i=1}^n l_\theta(X_i),
\qquad  l_\theta(x) = \log f_\theta(x)
\end{equation}
For the GPD, this maximizer has no closed-form solutions and has to
be determined numerically, using a suitable initialization;
in our simulation study, we use the {\rm Hybr} estimator defined below.
\medskip

%--------------------------------------------------------------------------------
Next, we discuss the optimally-robust estimators. By Proposition~\ref{scalinvprop} all
of them achieve scale-invariance respectively shift-invariance passing to the log-scale as
in \eqref{logtrafo1}, and all of them use a one-step construction \eqref{onestep}
with {\rm Hybr} as starting estimator.

%--------------------------------------------------------------------------------
\mparagraph{MBRE}
%--------------------------------------------------------------------------------
%
Minimizing the maximal bias on convex contamination neighborhoods,
we obtain the MBRE estimator, see Proposition~\ref{OptRobProp}; in the terminology of \citet{Ha:Ro:86}
this is the \textit{most B-robust} estimator. In most references though, e.g.\ \citet{Du:98},
one uses M-equations instead of one-step constructions to achieve  IF  $\bar \psi$ from
Proposition~\ref{OptRobProp}. %\ACHTUNG{TRUE?}
At $\xi=0.7$ and $\beta=1$, we obtain the following Lagrange multipliers $A$, $a$, $b$
\begin{eqnarray}
\nquad\nquad A_{\ssr MBRE}&=& \left(
\begin{array}{rr}
 1.00, &-0.18\\
-0.18, & 0.22
\end{array}\right),
\quad
a_{\ssr MBRE}= (-0.18, 0.00),\quad
b_{\ssr MBRE}= 3.67
\end{eqnarray}
$b_{\ssr MBRE}$ is unique while $A_{\ssr MBRE}$ and $a_{\ssr MBRE}$ are only unique up to
a scalar factor, which in our context is fixed setting $A_{1,1}=1$.

%
%--------------------------------------------------------------------------------
\mparagraph{OMSE}
%--------------------------------------------------------------------------------
%
 For OMSE we proceed similarly as for MBRE, i.e., we determine $\hat \psi$
 according to Proposition~\ref{OptRobProp}.
%There are again no closed form expressions for $A$, $a$, and $b$, but
%corresponding algorithms to determine $A$, $a$, and $b$ are implemented
%to {\sf R} within the {\tt ROptEst} package available on CRAN.
At $\xi=0.7$ and $\beta=1$,  we obtain the unique
Lagrange multipliers
\begin{eqnarray}
\nquad\nquad A_{\ssr OMSE}&=\left(
\begin{array}{rr}
 10.26,& -2.89 \\
-2.89, & 3.87
\end{array} \right)
,\quad
a_{\ssr OMSE}= (-1.08, 0.12),\quad
b_{\ssr OMSE}&=4.40
\end{eqnarray}
%
%The use of norm $n_\beta$ enforces (asympt.) in-/equivariance,
%i.e.; \eqref{eqMBRE} holds mutatis mutandis,
%
%
\begin{Remark}
OMSE also solves the
``Lemma~5 problem'' with its own GES as bias bound, compare \citep[Thm.~5.5.7]{Ried:94}, i.e.,
among all ALEs minimizes the (trace of the) asymptotic variance
subject to this bias bound on neighborhood ${\cal U}_n$.
Hence OMSE is a particular OBRE in the terminology of \citet{Ha:Ro:86},
spelt out for the GPD case in \citet{Du:98} (but again using M equations instead
of a one-step construction).
She does not head for the MSE-optimal bias bound, so our OMSE
will in general be better than her OBRE w.r.t.\ MSE at radius $r$.
On the other hand, for given a bias bound $b$, equations~\eqref{hatpsidef} also
yield a radius $r(b)$ for which a given OBRE is MSE-optimal. In this sense,
bias bound $b$ and radius $r$ are equivalent parametrizations of
degree of robustness required for the solution.
\end{Remark}
%

%--------------------------------------------------------------------------------
\mparagraph{RMXE}
%--------------------------------------------------------------------------------
%
 As mentioned, the RMXE is obtained by maximizing ${\rm eff.ru}$ among all ALEs $S_n$.
 By \citet[Thm.~6.1]{R:R:04}, we have
 \begin{equation} \label{eqeffru}
 {\rm eff.ru}(S_n)=\min\big({\rm eff.id}(S_n),
{\rm GES}^2({\rm MBRE})/{\rm GES}^2(S_n)\big)
 \end{equation} %which for fixed $g:={\rm GES}(S_n)$ is
%maximized by the respective ${\rm OBRE}$ with bias bound $g$. So for
and the RMXE is the OBRE with GES $b$ equalling both terms in the $\min$-expression in
\eqref{eqeffru}.
%Again, corresponding algorithms to determine $A$, $a$, and $b$ are implemented
%to {\sf R} within the {\tt ROptEst} package available on CRAN.
In our model at $\xi=0.7$ and $\beta=1$,  we obtain
the unique Lagrange multipliers

\begin{eqnarray}
\nquad\nquad A_{\ssr RMXE}&=\left(
\begin{array}{rr}
 10.02,& -2.87 \\
-2.87, &  3.85
\end{array} \right)
,\quad
a_{\ssr RMXE}= (-1.03, 0.12),\quad
b_{\ssr RMXE}&=4.44
\end{eqnarray}

\begin{Remark}
Passing from MSE to another risk does not in general invalidate our optimality, compare
\citet[Thm.~3.1]{R:R:04}. Whenever the asymptotic risk is representable
as $G(\tr \,\asVar, |\asBias|)$ for some function $G$ isotone in
both arguments, the optimal $\IF$ is again
in the class of OBRE estimators---with possibly another bias weight.
In addition, the RMXE for MSE is simultaneously optimal for all
homogenous risks of this form with continuous $G$ (Thm.~6.1 loc.cit.). In particular, for one-dimensional parameter,
this covers all risks of type  $\Ew |S_n-\theta|^p$ for any $p \in [1,\infty)$.
\end{Remark}

%We continue with.

%----------------------------------------------------------------------------------
\subsection{Starting Estimators}
%----------------------------------------------------------------------------------
Initializations for the estimators discussed so far are provided
by the next group of estimators (PE, MMed, MedkMAD, Hybr). They
can all be shown to fulfill the requirements given in
\citet[Ch.~6]{Ried:94}, in particular they are uniformly $\sqrt{n}$-tight
on our shrinking neighborhoods. Corresponding proofs are available
upon request.

%--------------------------------------------------------------------------------
\mparagraph{PE} %\label{PEsec}
%--------------------------------------------------------------------------------
%
Estimators based on the empirical quantiles of GPD are described
in the Elementary Percentile Method (EPM) by \citet{C:H:97}.
Pickands' estimator (PE), a special case of EPM,  is based on the
empirical 50\%  and 75\%  quantiles $\hat Q_2$ and $\hat Q_3$ respectively,
and has first been proposed by \citet{Pick:75}.
The construction behind PE is not limited to $50\%$ and $75\%$ quantiles.
More specifically, let $a>1$ and consider the empirical $\alpha_i$-quantiles
for $\alpha_1 = 1-1/a$ and $\alpha_2 = 1-1/a^2$
denoted by $\hat Q_2(a)$,
$\hat Q_3(a)$, respectively. Then PE is obtained
for $a=2$, and as theoretical quantiles we obtain
$Q_2(a) = \frac{\beta}{\xi}(a^\xi - 1)$,
$Q_3(a) = \frac{\beta}{\xi}(a^{2\xi} - 1)$,
and the (generalized) PE denoted by PE(a) for $\xi$ and $\beta$ is
\begin{equation}
\hat{\xi} = \Tfrac{1}{\log a}\log \Tfrac{\hat Q_3(a) - \hat Q_2(a)}{\hat Q_2(a)},
\quad \hat{\beta} = \hat{\xi}  \Tfrac{{\hat Q_2(a)}^2}{\hat Q_3(a)-2\hat Q_2(a)}
\end{equation}
%%

%--------------------------------------------------------------------------------
\mparagraph{MMed} %\label{MMedsec}
%--------------------------------------------------------------------------------
%
The method of medians estimator of \citet{P:W:01} consists of fitting
the (population) medians of the two coordinates of the score function
$\Lambda_\theta$ against the corresponding sample medians of $\Lambda_\theta$, i.e.;
we have to solve the system of equations
\begin{align}
{\rm median}(X_i)/\beta=m_\xi,\qquad \mbox{for } m_\xi:=F_{1,\xi}^{-1}(1/2)=(2^\xi-1)/\xi\\
{\rm median}\Big(\log(1+\xi X_i/\beta)\beta^{-2}-%
 (1+\xi)X_i(\beta\xi+\xi^2X_i)^{-1}\Big)=M(\xi)
\end{align}
where $M(\xi)$ is the population median of the $\xi$-coordinate
of $\Lambda_{\theta_1}(X)$ with $X\sim{\rm GPD}(\theta_1)$.
Solving the first equation for $\beta$ and plugging in the corresponding
expression into the second equation, we obtain a one-dimensional
root-finding problem to be solved, e.g.\ in {\sf R} by {\tt uniroot}.

%--------------------------------------------------------------------------------
\mparagraph{MedkMAD} %\label{Sec:MedkMAD}
%--------------------------------------------------------------------------------
%
Instead of matching empirical moments
against their model counterparts, an alternative is to match corresponding
location and dispersion measures; this gives
\textbf{L}ocation-\textbf{D}ispersion estimators, introduced by \citet{Ma:99}.
While a natural candidate for the location part is given by the median,
for the dispersion measure, promising candidates are given by
the median of absolute deviations MAD and the alternatives Qn and Sn
introduced in \citet{Ro:93}, producing estimators MedMAD, MedQn, and MedSn,
respectively. All these pairs are well known for
their high breakdown point in location-scale models, jointly
attaining the highest possible ABP of $50\%$ 
among all affine equivariant
estimators at symmetric, continuous univariate distributions.
For results on MedQn and MedSn, see \citet{Ru:Ho:10b}. These results
justify our restriction to Med(k)MAD for the GPD model in this paper.

Due to the considerable skewness to the right of the GPD, MedMAD
can be improved by using a dispersion measure that takes this skewness
into account. For a distribution $F$ on $\R$ with median $m$
let us define for $k>0$
\begin{equation}
{\rm kMAD}(F,k):=\inf\big\{\,t>0\,\big|\, F(m+kt)-F(m-t)\ge 1/2\,\big\}
\end{equation}
where $k$ in our case is chosen to be a suitable number
larger than $1$, and $k=1$ would reproduce the MAD. Within the
class of intervals about the median $m$ with covering probability $50\%$,
we only search those where the part right to $m$ is $k$ times longer
than the one left to $m$. Whenever $F$ is continuous, kMAD preserves
the FSBP of the MAD of $50\%$.
The corresponding estimator for $\xi$ and $\beta$ is called
\textit{MedkMAD} and consists of two estimating equations.
The first equation is for the median of the GPD, which is
$m =m(\xi,\beta)= \beta(2^\xi-1)/\xi $.
The second equation is for the respective kMAD, which has
to be solved numerically as unique root $M$ of $f_{m,\xi,\beta;k}(M)$  for
\begin{equation} \label{kMADdef}
f_{m,\xi,\beta;k}(M)=1/2+ \tilde v_{m,M,\xi,\beta}(k) - \tilde v_{m,M,\xi,\beta}(-1)
\end{equation}
where
$%\begin{equation}
\tilde v_{m,M,\xi,\beta}(s):=(1+\xi(sM+m)/\beta)^{-1/\xi} %\label{M+-def}
$. %\end{equation}

%%+++++++++++++++++++++++++++++++++++++++++++++++++++
\mparagraph{Hybr}
%%+++++++++++++++++++++++++++++++++++++++++++++++++++
Still,  Table~\ref{Tab.n40} here and Table~9 of \citet{Ru:Ho:10}
show failure rates of  $8\%$ for $n=40$ and $2.3\%$ for $n=100$
to  solve the MedkMAD equations for $k=10$.
To lower these rates we propose a hybrid
estimator {\rm Hybr}, that by default returns MedkMAD for $k=10$,
and by failure tries several $k$-values in a loop (at most $20$)
returning the first estimator not failing. We start at $k=3.23$
(producing maximal ABP), and at each iteration multiply $k$ by $3$. This leads to
failure rates of $2.3\%$ for $n=40$ and $0.0\%$ for $n=100$.
Asymptotically, {\rm Hybr} coincides with MedkMAD, $k=10$.

%--------------------------------------------------------------------------------
\subsection{Competitor Estimators}
%--------------------------------------------------------------------------------
The following estimators were suggested to us in an application to operational risk, see \citet{Ru:Ho:10}.
%no SMLE or MDE in this reference:,see \citet{R:H:B:11}.

\mparagraph{SMLE} %\label{SMLEsec}
%--------------------------------------------------------------------------------
%
Skipped Maximum Likelihood Estimators (SMLE) are ordinary MLEs,
skipping the largest $k$
observations. This has to be distinguished from the better investigated
\textit{trimmed/weighted MLE}, studied by \citet{F:S:94}, \citet{H:L:97},
\citet{V:N:98}, \citet{M:N:01}, %\linebreak[4]
where trimming/weighting is done according
to the size (in absolute value) of the log-likelihood.\\
In general these concepts fall apart as they refer to different orderings;
in our situation they coincide due to the monotonicity of the
likelihood in the observations.

As this skipping is not done symmetrically, it induces a non-vanishing
bias $B_n=B_{n,\theta}$ already present in the ideal model.
To cope with such biases three strategies can be used---the first two
already considered in detail in \citet[Section~2.2]{D:M:02}:
(1) correcting the criterion function for the skipped summands, (2)
correcting the estimator for  bias $B_n$, and (3)
no bias correction at all, but, conformal to our
shrinking neighborhood setting, to let the skipping proportion
$\alpha$  shrink at the same rate. Strategy~(3) reflects
the common practice where $\alpha$ is often chosen small, and the
bias correction is omitted. In the sequel, we only study Strategy~(3)
with $\alpha=\alpha_n=r'/\,\sqrt{n\,}$ for some $r'$ larger than
the actual $r$. This way indeed bias becomes asymptotically negligible: %,
%as shown in the following lemma. % a proof of which is contained in
%\citet[Lem.~2.1]{Ru:Ho:10}. %\vspace{-2ex}

\begin{lemma} \label{Lem2.1}
In our ideal GPD model, the bias $B_n$ of SMLE
with skipping rate $\alpha_n$ is bounded from above by
$\bar c \alpha_n \log(n)$ for some $\bar c<\infty$, eventually in $n$.

If for some $\zeta\in (0,1]$, $\liminf_n\alpha_n n^\zeta > 0$,
then for some $\underline{c}>0$ also\newline
\mbox{\hspace{1cm}}$\liminf_n n^\zeta B_n \ge \underline{c} \liminf_n n^\zeta\alpha_n \log(n)$.

If $0 < \underline{\alpha}=\liminf_n\alpha_n <\alpha_0$
for $\alpha_0=\exp(-3-1/\xi)$, then for some $\underline{c}'>0$\newline
\mbox{\hspace{1cm}}$\liminf_n B_n \ge \underline{c}' \underline{\alpha} (-\log(\underline{\alpha}))$.
\end{lemma}

It can be shown along the lines of \citet[Thm.~1.6.6]{Ried:94}
that after subtracting bias $B_n$, SMLE is indeed an ALE.

%\noindent Hence, the we need to correct
%for the then considerable bias. %\medskip

%In view of \citet{Ruck:10}, for $\alpha_n=r'/\sqrt{n}$, this makes for an
%admissible starting estimator. Yet,
%----------------------------------------------------------------------------------
%\subsection{MDE-type Estimators}
%----------------------------------------------------------------------------------
%--------------------------------------------------------------------------------
\mparagraph{MDE}
%--------------------------------------------------------------------------------
%
General  minimum distance estimators (MDEs) are defined as minimizers of a
suitable distance between the theoretical $F$ and empirical
distribution $\hat{F}_n$.
Optimization of this distance in general has to be done
numerically and, as for MLE and SMLE, depends on a
suitable initialization (here again:  {\rm Hybr}). We use Cram\'er-von-Mises distance
defined for c.d.f.'s $F$, $G$ and some $\sigma$-finite measure $\nu$ on $\B^k$ as
\begin{equation}
 d_{\ssr CvM}(F,G)^2= \int (F(x)-G(x))^2 \, \nu(dx)
\end{equation}
i.e.; %
%
%\begin{equation}
${\rm MDE} = \argmin\nolimits_{\theta} d_{\ssr CvM}(\hat F_n,F_\theta)
$.
%\end{equation}
%
In this paper we use $\nu=F_\theta$.
Another common setting in the literature uses the empirical, $\nu=\hat F_n$.
%The {\rm Hybr} estimator defined below again serves as initialization.
%As initialization we again use  estimator defined below.
As shown in \citet[Ex.~4.2.15, Sec~6.3.2]{Ried:94}, CvM-MDE belongs to the
class of ALEs.
%---------------------------------------------------------------------
\subsection{Computational and Numerical Aspects}\label{CompNumAspects}
%---------------------------------------------------------------------
%\vspace{-.5ex}

\noindent For computations, we use {\sf R} packages of \citet{R}, and
addon-packages {\tt ROptEst}, \citet{ROptEst} and
\texttt{POT}, \citet{Rib:09}, available on the \textit{Comprehensive
R Archive Network} {\tt CRAN},
\url{cran.r-project.org}. %{{\tt cran.r-project.org}}.

%Our estimators, as to computation, can be divided into four classes:%
%
%1. Estimators in closed-form expressions like PE
%(after possibly sorting the observations).
%As to computation time, their evaluation is by magnitudes faster
%than of the other groups, which makes them attractive for batch uses.
%
%2. M-estimators like MLE, SMLE, and MDE, obtained by optimizing
%a corresponding criterion function and solved iteratively
%by using {\sf R} function {\tt optim} and hence need a suitable
%initialization to find the ``right'' local optimum.
%
%3. Z-estimators like MMed and MedkMAD, i.e.; the zero of a(n)
%(system of) equation(s). In fact, both cases may be
%reduced to univariate problems, hence may use {\sf R}
%function {\tt uniroot}, with canonical search interval.
%
%4. One-step constructions like MBRE, OMSE, and RMXE, depending
%on a suitably chosen starting estimator. Once this starting estimate
%is found and the respective influence function at the starting
%estimate determined, computation of MBRE, OMSE, and RMXE
%is extremely fast, just involving an average.\\
%..........................................
\mparagraph{Computation of Lagrange multipliers} %\label{compasp}
%..........................................
 $A$, $a$, and $b$ of the optimally-robust
IFs from Proposition~\ref{OptRobProp} (at the starting estimate)
 are not available in closed form expressions,
but corresponding algorithms  to determine them for each of MBRE, OMSE,
and RMXE are implemented in {\sf R} within
 package {\tt ROptEst} \citep{ROptEst} available on CRAN.
Although these algorithms cover general $L_2$-differentiable models,
particular extensions are needed for the computation of the
expectations under the heavy-tailed GPD.

%\smallskip

%..........................................
\mparagraph{Speed-up by interpolation} \label{compasp}
%..........................................
Due to the lack of invariance in $\xi$, solving for
equations~\eqref{barpsidef} and \eqref{hatpsidef} can be quite slow:
for any starting estimate  the solution
has to be computed anew. Of course, we can reduce the problem by one
dimension due to Proposition~\ref{scalinvprop}, i.e.;
we only would need to know the influence functions for ``all'' values
$\xi>0$. To speed up computation, % especially for our simulation study,
we therefore have used the following approximative approach, already
realized in M.~Kohl's {\sf R} package {\tt RobLox}  \citep{RobLox} for
the Gaussian one-dimensional location and scale model\footnote{Due
to the affine equivariance of MBRE, OBRE, OMSE in the location and
scale setting, interpolation in package {\tt RobLox} is done only
 for varying radius $r$.}.
In our context, the speed gain obtainable by Algorithm~\ref{algor} is by a factor
of  $\sim 125$, and for larger $n$ can be increased by yet another
factor $10$ if we skip the re-centering/stan\-dar\-di\-zation and instead
return $Y^\natural w^\natural$.

\begin{Algorithm}
\label{algor}
%\sf
For a grid $\xi_1,\ldots, \xi_M$ of values
of $\xi$, giving parameter values $\theta_{i,1}=(\xi_i,1)$ (and for OMSE
to given $r=0.5$), we offline determine
the optimal $\IF$'s $\psi_{\theta_{i,1}}$, solving
equations~\eqref{barpsidef} and \eqref{hatpsidef} for each $\theta_{i,1}$
and store the respective Lagrange
multipliers $A$, $a$, and $b$, denoted by
$A_i$, $a_i$, $b_i$.
In the evaluation of the ALE for given
starting estimate $\theta^{(0)}_n$, we use Proposition~\ref{scalinvprop}
and pass over to parameter value
$\theta'=(\xi^{(0)}_n,1)$. For  $\theta'$, we find values
$A^\natural$, $a^\natural$, and $b^\natural$ by
interpolation for the stored grid values
$A_i$, $a_i$, $b_i$. This gives us
$Y^\natural=A^\natural \Lambda_{\theta'}- a^\natural$,
and $w^\natural = \min\big(1, b^\natural/|Y^\natural|)\big)$.
So far, $Y^\natural w^\natural\not\in \Psi_2(\theta')$, i.e., does
not satisfy \eqref{ICcond} at $\theta'$. Thus, similarly to \citet[Rem.~5.5.2]{Ried:94}, we
define  $Y^\sharp=A^\sharp \Lambda_{\theta'}-a^\sharp$
for $a^\sharp = A^\sharp z^\sharp$,
$%\begin{equation}
z^\sharp=\Ew_{\theta'}[\Lambda_{\theta'}w^\natural]/%
\Ew_{\theta'}[w^\natural],\quad %
A^\sharp=\big\{\Ew_{\theta'}[(\Lambda_{\theta'}-z^\sharp)
(\Lambda_{\theta'}-z^\sharp)^\tau w^\natural]\big\}^{-1},
$ %\end{equation}
and pass over to  $\psi^\sharp=Y^\sharp w^\natural$.
By construction $\psi^\sharp\in\Psi_2(\theta')$.
\end{Algorithm}
%\begin{Rem}
%(a) $\psi^\sharp$ produced in this way in general does not
%solve \eqref{barpsidef} and \eqref{hatpsidef}, i.e.\
%$A^\natural\not=A^\sharp$, $a^\natural\not=a^\sharp$, nor holds
%$b^\natural\not=b^\sharp$, but if the
%grid is dense enough, due to the smoothness
%of our model, we will have approximate equality in all
%these equations. For this smoothness see \citet[Figure~2]{Ru:Ho:10}.
%
%We have checked the accuracy in terms of efficiency
%loss w.r.t.\ the actual optimal $\IF$ in terms of relative
%$\asMSE$. At the true parameter $\xi=0.7$, our computations give $99.3\%$ efficiency
%for OMSE and $99.0\%$ for MBRE, while at $\xi=0.1$, $\xi=1.3$ we never drop
%below $99\%$ efficiency.\\
%\noindent(b)
%\end{Rem}
%
%We apply Algorithm~\ref{algor} for both MBRE and OMSE.

%\vspace{-2.5ex}

%-------------------------------------------------------------------------------------------------------
\subsection{Synopsis of the Theoretical Properties}\label{Synopsis}
%-------------------------------------------------------------------------------------------------------
%
%..........................................
\mparagraph{Breakdown, bias, variance, and efficiencies:}
%..........................................
%
In Table~\ref{Tab1}, we summarize our
findings, evaluating criteria  ${\rm FSBP}$ (where exact values are available),
$\asBias=r \,{\rm GES}$, $\tr\, \asVar$, and $\asMSE$ (at $r=0.5$). To be
able to compare the results for different sample sizes $n$,
these figures are standardized by sample size $n$, respectively by $\sqrt{n}$ for the bias.
We also determine efficiencies  ${\rm eff.id}$, ${\rm eff.re}$, and ${\rm eff.ru}$.
For ${\rm FSBP}$ of MLE, SMLE, we evaluate terms at  $n=1000$,
where for SMLE we set $r'=0.7$ entailing $\alpha_n=2.2\%$.
Finally, we document the ranges of least favorable $x$-values $x_{\ssr l.f.}$,
at which the considered $\IF$s attain their GES.
These are the most vulnerable points of the respectively estimators infinitesimally,
as contamination therein will render bias maximal.
In all situations where  $x_{\ssr l.f.}$ is unbounded, a value $10^{10}$ will suffice
to produce maximal bias in the displayed accuracy. On the other hand, PE and MMed are most harmfully contaminated by
smallish values of about $x=1.5$ (for $\beta=1$).

The results for SMLE are to be
read with care: $\asBias$ and $\asMSE$ do not account for
the bias $B_n$ already present in the ideal model, but only for
the extra bias induced by contamination. Lemma~\ref{Lem2.1} entails that
$B_n$ is of exact unstandardized order $\LO(\log(n)/\sqrt{n})$,
hence, $\asBias$ and $\asMSE$ should both be infinite,
and efficiencies in ideal and contaminated situation be
$0$. For $n=1000$, $\asBias$ and $\asMSE$
are finite: according to Lemma~\ref{Lem2.1}, $\sqrt{1000\,}\,B_{1000}\approx 5.38$,
while the entry of $3.75$ in Table~\ref{Tab1} is just ${\rm GES}$.
%and is at large due to an underestimation of $\xi$ by $0.17$.

As noted, MLE achieves
smallest $\asVar$, hence is best in the ideal model, but
at the price of a minimal FSBP and an infinite GES, so at any sample
one large observation size suffices to render MSE arbitrarily large.

MedkMAD gives very convincing results in both asMSE and (E)FSBP.
It qualifies as a starting estimator, as it uses univariate root-finders with
parameter-independent search intervals.
The best breakdown behavior so far has been achieved by Hybr, with
$\ve^\ast\approx 1/3$ for a reasonable range of $\xi$-values.
MDE shares an excellent reliability with Hybr, but contrary
to the former needs a reliable starting value for the optimization.
%Its computation is quite fast, however.

%
MBRE, OMSE, and RMXE have bounded IFs and are constructed as
one-step estimators, so
by Lemma~\ref{Lemma_exp_scale} inherit the FSBP of the starting
estimator (Hybr), while at the same time MBRE achieves lowest GES
(unstandardized by $n$ of order
$0.1$ at $n=1000$), OMSE is best according to $\asMSE$, and
RMXE is best as to ${\rm eff.ru}$. RMXE (which is the OMSE for $r=0.486$) and OMSE for $r=0.5$,
with their radii almost coinciding, are virtually indistinguishable,  guaranteeing an efficiency
of $68\%$ over all radii.

We admit that MDE, MedkMAD/Hybr, and MBRE
are close competitors in both efficiency and FSBP,
both at given radius $r=0.5$ and
%
%Considering unknown contamination radius and least favorable
%efficiency ${\rm eff.ru}$, OMSE (RMXE) for $r=0.5$
%is best among the considered estimators and .
%MDE, MedkMAD/Hybr, and MBRE also give acceptable
as to their least favorable efficiencies,
never dropping considerably below $0.5$. All other estimators are
less convincing.
\begin{table}[ht]
\begin{center}
%\hspace{-3cm}
\begin{footnotesize}\begin{tabular}{l|rrrrrrcr}
  \hline
estimator&  $\!\!\!\asBias\!\!\!$ & $\!\!\!\tr\,\asVar$ & $\!\!\!\asMSE\!\!\!$ %
& {\small$\!\!{\rm  eff.id}\!\!$}  & {\small$\!\!{\rm eff.re}\!\!$} %
& {\small$\!\!\!{\rm eff.ru}\!\!\!$} &$\!\!\!x_{\ssr l.f.}\!\!\!$ %
&$\!\!\!\!\!\bar\ve^\ast_{1000}$ \\
  \hline
MLE    & $\infty  $& $ 6.29$ & $\infty $  & $1.00$ & $0.00$ & $0.00$ &  $\!\!\!\!\!\infty$                        \!\!\!\!\!\! & $\!\!0.00\hphantom{{}^{{\rm ?}}}   \!\!\!$\\
MBRE   & $   1.84 $& $13.44$ & $ 16.80$   & $0.47$ & $0.84$ & $0.47$ &  $\!\!\!\!\![0.00;\infty)$                 \!\!\!\!\!\! & $\!\!0.35^\ast                     \!\!\!$\\
OMSE   & $   2.20 $& $ 9.29$ & $ 14.13$   & $0.68$ & $1.00$ & $0.68$ &  $\!\!\!\!\![0.00;0.07]\cup[5.92;\infty)$  \!\!\!\!\!\! & $\!\!0.35^\ast                     \!\!\!$\\
${\rm RMXE}$   & $   2.22 $& $ 9.21 $ & $ 14.14$   & $ 0.68$ & $1.00$ & $0.68$ &  $\!\!\!\!\![0.00;0.07]\cup[5.92;\infty)$  \!\!\!\!\!\! & $\!\!0.35^\ast                     \!\!\!$\\
PE     & $   4.08 $& $24.24$ & $ 40.87$   & $0.26$ & $0.35$ & $0.20$ &  $\!\!\!\!\![0.89;2.34]$                   \!\!\!\!\!\! & $\!\!0.06\hphantom{{}^{{\rm ?}}}   \!\!\!$\\
MMed   & $   2.62 $& $17.45$ & $ 24.32$   & $0.36$ & $0.58$ & $0.32$ &  $\!\!\!\!\![0.00;0.34]\cup[0.90;2.54]$    \!\!\!\!\!\! & $\!\!0.25^{{\rm ?}}                \!\!\!$\\
MedkMAD& $   2.19 $& $12.80$ & $ 17.60$   & $0.49$ & $0.80$ & $0.49$ &  $\!\!\!\!\![0.54;0.89]\cup[4.42;\infty)$  \!\!\!\!\!\! & $\!\!0.31\hphantom{{}^{{\rm ?}}}   \!\!\!$\\
SMLE   & $   3.75 $& $ 7.03$ & $ 21.08$   & $0.90$ & $0.67$ & $0.03$ &  $\!\!\!\!\![20.67;\infty)$                \!\!\!\!\!\! & $\!\!0.02\hphantom{{}^{{\rm ?}}}   \!\!\!$\\
MDE    & $   2.45 $& $ 9.76$ & $ 15.74$   & $0.64$ & $0.90$ & $0.56$ &  $\!\!\!\!\!\{0,\infty\}$                  \!\!\!\!\!\! & $\!\!0.35^{{\rm ?}}                \!\!\!$\\
 \hline
\end{tabular}\end{footnotesize}
\caption[Comparison of the asymptotic robustness properties of
the estimators]{\label{Tab1}{
Comparison of the asymptotic robustness properties of the estimators\\
${\hphantom{0}}^{\ast}$:
inherited from starting estimator ${\rm Hybr}$;
${\hphantom{0}}^{{\rm ?}}$:
conjectured.
}}
\end{center}
\uP
\end{table}
%\smallskip

\begin{figure}[vt]
\uQi
\centering
\includegraphics[width = 1\textwidth,height=11cm]{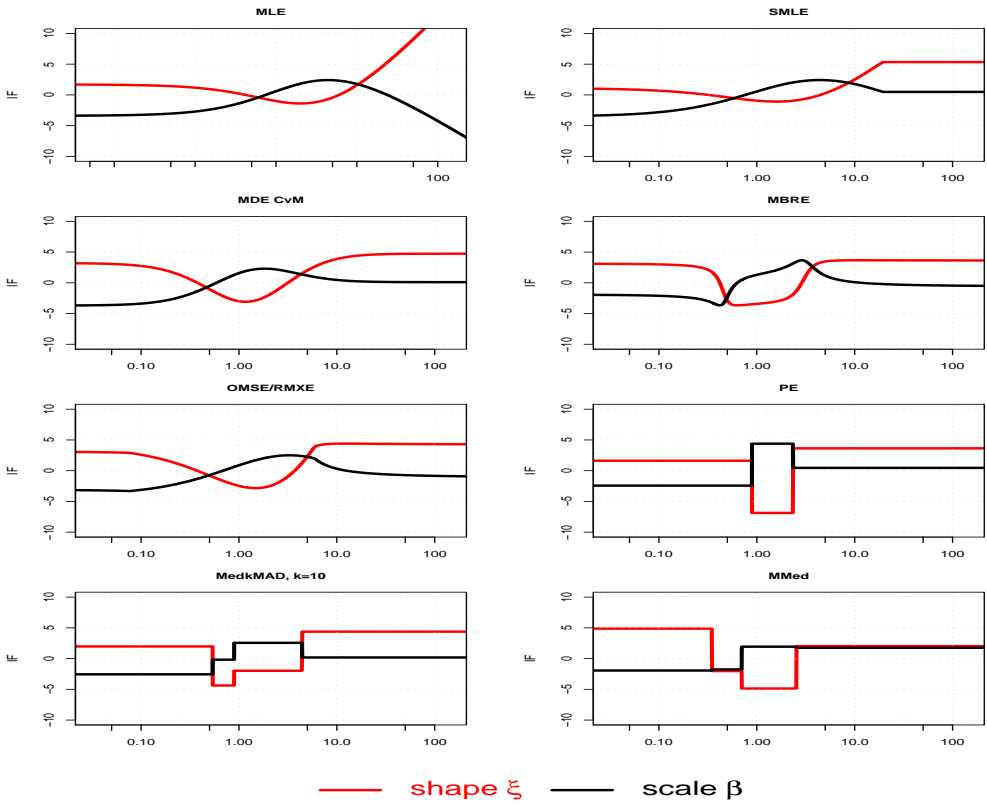}
\caption[Influence Functions of Estimators]{\label{Fig:ICs}{Influence Functions
of MLE, SMLE (with  $\approx0.7\cdot\sqrt{n}$
skipped value), MDE CvM, MBRE, OMSE, PE,
MMed, MedkMAD   estimators of the generalized Pareto distribution; mind the logarithmic scale of the $x$-axis.
}}
\uPi
\end{figure}

%..........................................
\mparagraph{Influence functions:}
%..........................................
In Figure~\ref{Fig:ICs},
we display the $\IF$s $\psi_\theta$ of the considered estimators.
The $\IF$ of RMXE  visually coincides
with the one of OMSE. All $\IF$s are scale invariant so that
$\psi_{\theta}(x) = d_\beta \psi_{\theta_1}(x/\beta)$.

Intuitively, based on optimality within $L_2(F_\theta)$,
to achieve high efficiency, the $\IF$ should be as close as
possible in $L_2$-sense to the respective optimal one.
So on first glance, MedkMAD achieves an astonishingly reasonable
efficiency in the contaminated situation, although its $\IF$ looks
quite different from the optimal one of OMSE; but, of course,
this difference occurs predominantly in regions of low $F_\theta$-probability.

%\medskip

\begin{figure}[vt]
\uQii
\centering
\includegraphics[width = 0.9\textwidth,height=13cm]{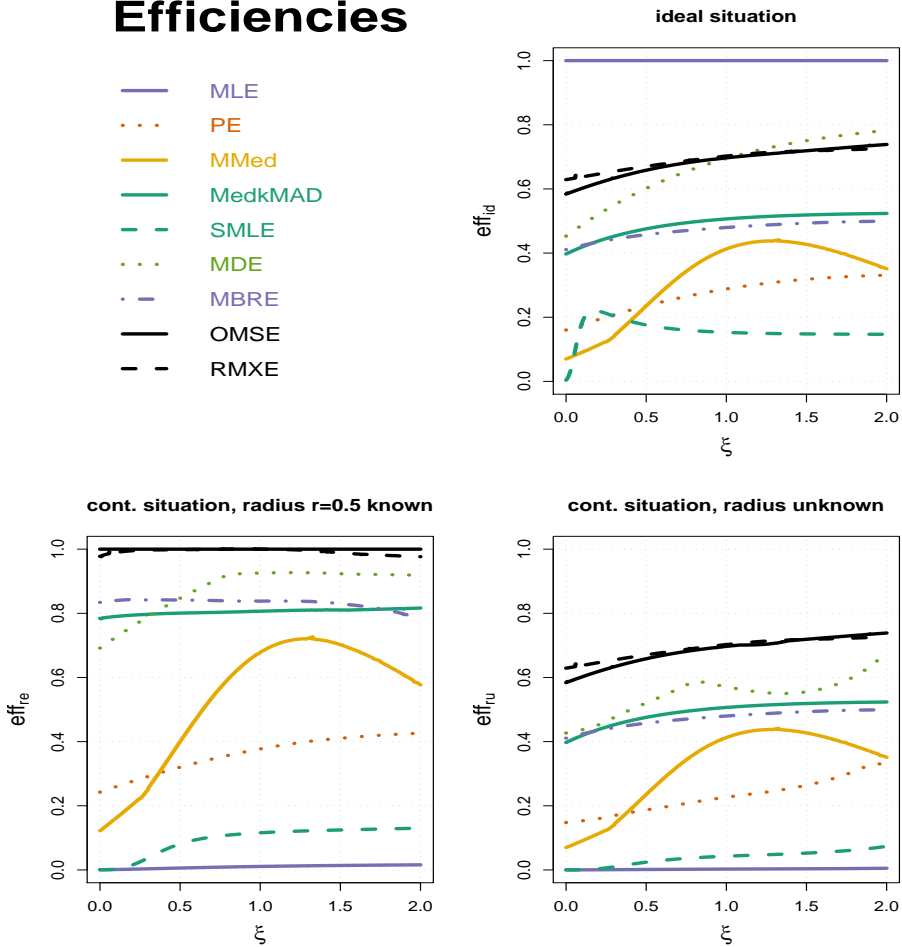}
\caption[Efficiencies for varying shape]{\label{xivarPic} {Efficiencies for varying shape
of MLE, SMLE (with  $\approx0.7\cdot\sqrt{n}$
skipped value), CvM-MDE, MBRE, OMSE, PE, MMed, MedkMAD
  estimators for scale $\beta=1$ and varying shape $\xi$. }}
\uPii
\end{figure}

%..........................................
\mparagraph{Values {\boldmath$\bf \xi\not=0.7$}:}
%..........................................
The behavior for our reference value $\xi=0.7$ is typical.
The conclusions we just have drawn as to obtainable efficiencies and
the ranking of the procedures largely remain valid for other parameter values,
as visible in Figure~\ref{xivarPic}. The least favorable radii for $\xi\in[0,2]$
all range in $[0.39, 0.51]$.
Note that due to the scale invariance we do not need to consider $\beta\not=1$.
From this figure we may in particular see the minimal value for the efficiencies
as extracted in Table~\ref{tabmin}. %The fact that the entry for RMXE is worse than
%the one for OMSE in the last line is

\begin{table}[ht]
\begin{center}
\begin{footnotesize}
\begin{tabular}{c|ccccccccc}
\hline
estimator& MLE &     PE  &   MMed  &  MedkMAD & SMLE &   MDE   &  MBRE&    OMSE&    RMXE\\
\hline
$\min_{\xi}{\rm eff.id}$ & $1.00$ & $0.16$ &  $0.07$ & $0.40$ & $0.00$ & $0.45$ & $0.41 $ & $0.58$& $0.63$\\
$\min_{\xi}{\rm eff.re}$ & $0.00$ & $0.24$ &  $0.12$ & $0.78$ & $0.00$ & $0.69$ & $0.78 $ & $1.00$& $0.98$\\
$\min_{\xi}{\rm eff.ru}$ & $0.00$ & $0.15$ &  $0.07$ & $0.40$ & $0.00$ & $0.43$ & $0.41 $ & $0.58$& $0.63$\\
\hline
\end{tabular}
\end{footnotesize}
\caption[Minimal efficiencies for varying shape]{\label{tabmin} %
{Minimal efficiencies  for $\xi$ varying in $[0,2]$ in the ideal
model and for contamination of known and unknown radius}}
\end{center}
\uP
\end{table}

\section{Simulation Study}\label{Simulation Study}
\subsection{Setup}
For sample size $n=40$, we simulate  data from both the ideal GPD
with parameter values $\mu=0$, $\xi=0.7$, $\beta=1$. Additional tables and plots
for $n=100,1000$ can be found in \citet{Ru:Ho:10}.
We evaluate the estimators from the previous section at $M=10000$ runs in the
 respective situation (ideal/contaminated).

The contaminated data stems from the (shrinking) Gross Error Model~\eqref{GEM},
\eqref{shrinkRate} with $r=0.5$. For $n=40$, this amounts an actual contamination
 rate of $r_{40}=7.9\%$.

In contrast to other approaches, for realistic comparisons we allow for
\textit{estimator-specific contamination}, such that each estimator has to prove its usefulness
in its \textit{individual worst contamination situation}. This is particularly important for
estimators with redescending IF like PE and MMed, where drastically large observations
will not be the worst situation to produce bias.
As contaminating data distribution, we use $G_{n,i}={\rm Dirac}(10^{10})$,
except for estimators PE and MMed, where we use
$G_{n,i}'={\rm unif}(1.42,1.59)$ in accordance with $x_{\ssr l.f.}$ from Table~\ref{Tab1}.
%
%For MMed and MedkMAD for maximal MSE we should use $G_{n,i}$,
%while $G_{n,i}'$ produces higher failure rates, so
%for all entries except for the failure rate, we use  $G_{n,i}$ and for column
%``{\rm NA}'' we use  $G_{n,i}'$.
%
\subsection{Results}
Results are summarized in Table~\ref{Tab.n40}.
Values for  ${\rm  Bias}$, $\tr \,{\rm Var}$, and MSE (standardized
by $\sqrt{40}$ and $40$, respectively) all come with corresponding
CLT-based $95\%$-confidence intervals.
Column ``${\rm NA}$'' gives the failure rate in the computation in percent;
basically, these are failures of {\rm MMed} or {\rm MedkMAD}/{\rm Hybr} to find a zero,
which due to the use of {\rm Hybr} as initialization are then propagated to
MLE, SMLE, MDE, MBRE, OMSE, and RMXE. Column ``time'' gives the aggregated
computation time in seconds on a recent dual core processor for the $10000$
evaluations of the estimator for ideal and contaminated situation.
For MLE, SMLE, MDE, MBRE, OMSE, and RMXE we do not include the time for evaluating
the starting estimator ({\rm Hybr}) but only mention the values for the evaluations
given the respective starting estimate.
The respective best estimator is printed in bold face.
\begin{table}
\begin{center}\begin{footnotesize}
\begin{tabular}{c|ccrrrrrrrcrr}
\multicolumn{11}{c}{ideal situation:}\\[1ex]
\hline
estimator  &%
\multicolumn{2}{c}{${\rm  |Bias|}$} &\multicolumn{2}{c}{$\tr\,{\rm Var}$} %
 &\multicolumn{2}{c}{${\rm MSE}$}&\multicolumn{1}{c}{${\rm eff.id}$}%
 &\multicolumn{1}{c}{${\rm rank}$}&\multicolumn{1}{c}{${\rm NA}$}&time\\%
\hline
% %%%%%%%%%%
 {\bf MLE      } &  {\boldmath$ \bf 0.55$} &  {\boldmath$\!\!\!\!{\SSs\pm \bf 0.05}$} &  {\boldmath$    \bf 7.41        $} &  {\boldmath$\!\!\!\!{\SSs\pm     \bf 0.21}$} &  {\boldmath$   \bf 7.72        $} &  {\boldmath$\!\!\!\!{\SSs\pm   \bf  0.21}$} &  {\boldmath$ \bf 1.00$} &  {\bf   1} &  {\boldmath$   \bf 0.53$} &  {\boldmath$ 113$}\\
 {    MBRE  } &   {         $ 0.61$} &  {         $\!\!\!\!{\SSs\pm  0.08}$} &  {         $   18.62        $} &  {         $\!\!\!\!{\SSs\pm     1.56}$} &  {         $   19.00        $} &  {         $\!\!\!\!{\SSs\pm     1.59}$} &  {         $  0.41$} &  {      7} &  {         $   0.53$} &  {         $ 402$}\\
 {    OMSE  } &  {         $ 0.25$} &  {         $\!\!\!\!{\SSs\pm  0.06}$} &  {         $    9.02        $} &  {         $\!\!\!\!{\SSs\pm     0.22}$} &  {         $    9.08        $} &  {         $\!\!\!\!{\SSs\pm     0.21}$} &  {         $  0.85$} &  {      2} &  {         $   0.53$} &  {         $ 783$}\\
 {    RMXE  } &  {         $ 0.21$} &  {         $\!\!\!\!{\SSs\pm  0.06}$} &  {         $    9.27        $} &  {         $\!\!\!\!{\SSs\pm     0.33}$} &  {         $    9.31        $} &  {         $\!\!\!\!{\SSs\pm     0.32}$} &  {         $  0.83$} &  {      3} &  {         $   0.53$} &  {         $ 769$}\\
 {    PE       } &   {         $ 0.85$} &  {         $\!\!\!\!{\SSs\pm  0.27}$} &  {         $   19.30        $} &  {         $\!\!\!\!{\SSs\pm     1.54}$} &  {         $   20.01        $} &  {         $\!\!\!\!{\SSs\pm     1.67}$} &  {         $  0.39$} &  {     8} &  {         $   0.00$} &  {         $  13$}\\
 {    MMed     } &   {         $ 8.91$} &  {         $\!\!\!\!{\SSs\pm  1.98}$} &  {         $ 1.02\,{\rm e} 5$} &  {         $\!\!\!\!{\SSs\pm  2423.14}$} &  {         $ 1.02\,{\rm e} 5$} &  {         $\!\!\!\!{\SSs\pm  2458.24}$} &  {         $  0.00$} &  {     10} &  {         $  10.50$} &  {         $ 168$}\\
% {    MedMad   } &   {         $ 1.32$} &  {         $\!\!\!\!{\SSs\pm  0.10}$} &  {         $   24.77        $} &  {         $\!\!\!\!{\SSs\pm     1.30}$} &  {         $   26.52        $} &  {         $\!\!\!\!{\SSs\pm     1.39}$} &  {         $  0.29$} &  {     9} &  {         $  20.70$} &  {         $ 150$}\\
 {    MedkMAD  } &  {         $ 0.47$} &  {         $\!\!\!\!{\SSs\pm  0.07}$} &  {         $   11.55        $} &  {         $\!\!\!\!{\SSs\pm     0.30}$} &  {         $   11.78        $} &  {         $\!\!\!\!{\SSs\pm     0.29}$} &  {         $  0.66$} &  {      5} &  {         $   8.15$} &  {         $ 197$}\\
 {    Hybr   } &   {         $ 0.71$} &  {         $\!\!\!\!{\SSs\pm  0.07}$} &  {         $   11.96        $} &  {         $\!\!\!\!{\SSs\pm     0.31}$} &  {         $   12.46        $} &  {         $\!\!\!\!{\SSs\pm     0.30}$} &  {         $  0.62$} &  {      6} &  {         $   0.53$} &  {         $ 223$}\\
 {    SMLE     } &   {         $ 4.70$} &  {         $\!\!\!\!{\SSs\pm  0.06}$} &  {         $    9.49        $} &  {         $\!\!\!\!{\SSs\pm     0.30}$} &  {         $   31.62        $} &  {         $\!\!\!\!{\SSs\pm     0.47}$} &  {         $  0.24$} &  {     9} &  {         $   0.53$} &  {         $  75$}\\
 {    MDE      } &    {         $ 0.40$} &  {         $\!\!\!\!{\SSs\pm  0.06}$} &  {         $   10.56        $} &  {         $\!\!\!\!{\SSs\pm     0.27}$} &  {         $   10.72        $} &  {         $\!\!\!\!{\SSs\pm     0.25}$} &  {         $  0.72$} &  {      4} &  {         $   0.53$} &  {         $ 384$}\\
%  {    OMSE.nc  } &    {         $ 0.93$} &  {         $\!\!\!\!{\SSs\pm  0.06}$} &  {         $   10.82        $} &  {         $\!\!\!\!{\SSs\pm     0.32}$} &  {         $   11.68        $} &  {         $\!\!\!\!{\SSs\pm     0.34}$} &  {         $  0.66$} &  {      5} &  {         $   0.53$} &  {         $  52$}\\
%  {    MBRE.nc  } &   {         $ 1.14$} &  {         $\!\!\!\!{\SSs\pm  0.09}$} &  {         $   23.48        $} &  {         $\!\!\!\!{\SSs\pm     1.11}$} &  {         $   24.77        $} &  {         $\!\!\!\!{\SSs\pm     1.20}$} &  {         $  0.31$} &  {     11} &  {         $   0.53$} &  {         $  48$}\\
%  {    RMXE.nc  } &    {         $ 0.90$} &  {         $\!\!\!\!{\SSs\pm  0.07}$} &  {         $   11.94        $} &  {         $\!\!\!\!{\SSs\pm     0.50}$} &  {         $   12.74        $} &  {         $\!\!\!\!{\SSs\pm     0.52}$} &  {         $  0.61$} &  {      8} &  {         $   0.53$} &  {         $  48$}\\
\hline
 \multicolumn{11}{c}{}\\[2ex]
\multicolumn{11}{c}{contaminated situation:}\\[1ex]
% %
 \hline
 estimator  &
 \multicolumn{2}{c}{${\rm  |Bias|}$} &\multicolumn{2}{c}{$\tr\,{\rm Var}$} %
  &\multicolumn{2}{c}{${\rm MSE}$}&\multicolumn{1}{c}{${\rm eff.re}$}%
  &\multicolumn{1}{c}{${\rm rank}$}&\multicolumn{1}{c}{${\rm NA}$}\\
% %%
\hline
 {    MLE      } &   {         $ 394.12$} &  {         $\!\!\!\!{\SSs\pm  22.92}$} &  {         $ 1.37\,{\rm e} 7$} &  {         $\!\!\!\!{\SSs\pm  1.20\,{\rm e} 6}$} &  {         $ 1.52\,{\rm e} 7$} &  {         $\!\!\!\!{\SSs\pm  1.37\,{\rm e} 6}$} &  {         $  0.00$} &  {     10} &  {         $   0.53$}\\
 {    MBRE  } &  {         $   1.70$} &  {         $\!\!\!\!{\SSs\pm   0.09}$} &  {         $   20.49        $} &  {         $\!\!\!\!{\SSs\pm     1.36        }$} &  {         $   23.37        $} &  {         $\!\!\!\!{\SSs\pm     1.39        }$} &  {         $  0.85$} &  {      4} &  {         $   0.37$}\\
 {    OMSE  } &    {         $   2.62$} &  {         $\!\!\!\!{\SSs\pm   0.07}$} &  {         $   13.11        $} &  {         $\!\!\!\!{\SSs\pm     0.42        }$} &  {         $   19.98        $} &  {         $\!\!\!\!{\SSs\pm     0.60        }$} &  {         $  0.99$} &  {      2} &  {         $   0.37$}\\
 {  \bf  RMXE  } &   {         $  \bf 2.73$} &  {         $\!\!\!\!{\SSs\pm   \bf 0.07}$} &  {         $   \bf 12.34        $} &  {         $\!\!\!\!{\SSs\pm    \bf  0.39        }$} &  {         $   \bf 19.80        $} &  {         $\!\!\!\!{\SSs\pm    \bf 0.57        }$} &  {         $ \bf 1.00$} &  {   \bf   1} &  {         $   0.37$}\\
 {    PE       } &   {         $   2.32$} &  {         $\!\!\!\!{\SSs\pm   0.49}$} &  {         $   62.25        $} &  {         $\!\!\!\!{\SSs\pm    67.90        }$} &  {         $   67.64        $} &  {         $\!\!\!\!{\SSs\pm    69.35        }$} &  {         $  0.30$} &  {     7} &  {         $   0.00$}\\
 {    MMed     } &   {         $   5.13$} &  {         $\!\!\!\!{\SSs\pm   1.17}$} &  {         $ 3563.54        $} &  {         $\!\!\!\!{\SSs\pm  1442.56        }$} &  {         $ 3589.87        $} &  {         $\!\!\!\!{\SSs\pm  1454.42        }$} &  {         $  0.01$} &  {     8} &  {         $   4.25$}\\
% {    MedMad   } &    {         $   1.01$} &  {         $\!\!\!\!{\SSs\pm   0.10}$} &  {         $   23.58        $} &  {         $\!\!\!\!{\SSs\pm     1.46        }$} &  {         $   24.61        $} &  {         $\!\!\!\!{\SSs\pm     1.44        }$} &  {         $  0.79$} &  {      7} &  {         $  37.49$}\\
 {    MedkMAD  } &   {         $   2.32$} &  {         $\!\!\!\!{\SSs\pm   0.09}$} &  {         $   18.82        $} &  {         $\!\!\!\!{\SSs\pm     0.49        }$} &  {         $   24.21        $} &  {         $\!\!\!\!{\SSs\pm     0.67        }$} &  {         $  0.82$} &  {      6} &  {         $   2.15$}\\
 {    Hybr   } &   {         $   2.23$} &  {         $\!\!\!\!{\SSs\pm   0.09}$} &  {         $   19.23        $} &  {         $\!\!\!\!{\SSs\pm     0.50        }$} &  {         $   24.21        $} &  {         $\!\!\!\!{\SSs\pm     0.67        }$} &  {         $  0.82$} &  {      5} &  {         $   0.02$}\\
 {    SMLE     } &  {         $   7.44$} &  {         $\!\!\!\!{\SSs\pm   3.10}$} &  {         $ 2.51\,{\rm e} 5$} &  {         $\!\!\!\!{\SSs\pm  1.52\,{\rm e} 5}$} &  {         $ 2.52\,{\rm e} 5$} &  {         $\!\!\!\!{\SSs\pm  1.52\,{\rm e} 5}$} &  {         $  0.00$} &  {     9} &  {         $   0.53$}\\
 {    MDE      } &   {         $   2.64$} &  {         $\!\!\!\!{\SSs\pm   0.08}$} &  {         $   16.19        $} &  {         $\!\!\!\!{\SSs\pm     0.43        }$} &  {         $   23.15        $} &  {         $\!\!\!\!{\SSs\pm     0.59        }$} &  {         $  0.86$} &  {      3} &  {         $   0.53$}\\
%  {    OMSE.nc  } &   {         $   2.34$} &  {         $\!\!\!\!{\SSs\pm   0.08}$} &  {         $   18.46        $} &  {         $\!\!\!\!{\SSs\pm     0.64        }$} &  {         $   23.93        $} &  {         $\!\!\!\!{\SSs\pm     0.80        }$} &  {         $  0.81$} &  {      6} &  {         $   0.37$}\\
%  {    MBRE.nc  } &   {         $   1.31$} &  {         $\!\!\!\!{\SSs\pm   0.10}$} &  {         $   23.58        $} &  {         $\!\!\!\!{\SSs\pm     1.06        }$} &  {         $   25.29        $} &  {         $\!\!\!\!{\SSs\pm     1.08        }$} &  {         $  0.77$} &  {     10} &  {         $   0.37$}\\
%  {\bf RMXE.nc  } &  {\boldmath$   \bf 2.05$} &  {\boldmath$\!\!\!\!{\SSs\pm  \bf 0.08}$} &  {\boldmath$  \bf 15.26        $} &  {\boldmath$\!\!\!\!{\SSs\pm    \bf  0.54        }$} &  {\boldmath$  \bf 19.46        $} &  {\boldmath$\!\!\!\!{\SSs\pm     \bf 0.61        }$} &  {\boldmath$  \bf 1.00$} &  {\bf   1} &  {\boldmath$   \bf 0.37$}\\
\hline
\end{tabular}\end{footnotesize}
\caption{\label{Tab.n40}
 Comparison of the empirical robustness properties of the estimators at sample size $n=40$ and with log-transformation \eqref{logtrafo1} used for the scale
 component; numbers in small print indicate CLT-based $95\%$ confidence intervals for the empirical values.}
 \end{center}
\uPTiii
 \end{table}

The simulation study confirms our findings of Section~\ref{Synopsis};
entries in Table~\ref{Tab.n40} follow the same pattern as the ones of
Table~\ref{Tab1}. This holds in particular for the ideal situation,
and for the efficiencies, where in the latter case Table~\ref{Tab1} provides
reasonable approximations already for $n=100$
\citep[Tables~8,9]{Ru:Ho:10}. %, with the exception for
%SMLE and the PE-variants.

The ranking given by asymptotics is essentially valid already at
sample size $40$---%
as predicted by asymptotic theory, ${\rm RMXE}$ and ${\rm OMSE}$ in their
interpolated and $\IF$-corrected variant $\psi^\sharp$ at significance $95\%$ are
the best considered estimator as to MSE, although MDE, ${\rm MBRE}$, and ${\rm Hybr}$
come close as to eff.re.

By using Hybr as starting estimator the number of failures can be kept low:
already at $n=40$, it is less than $1\%$ in the ideal model and about $3\%$
under contamination. This is not true for MMed and MedkMAD,
which suffer from up to $33\%$ failure rate at this $n$ under
contamination. So Hybr is a real improvement.

\begin{figure}[!htb]%[vt]
\uQiii
\centering
\subfigure[no contamination, sample size  $n=40$]{
\includegraphics[width=0.65\textwidth]{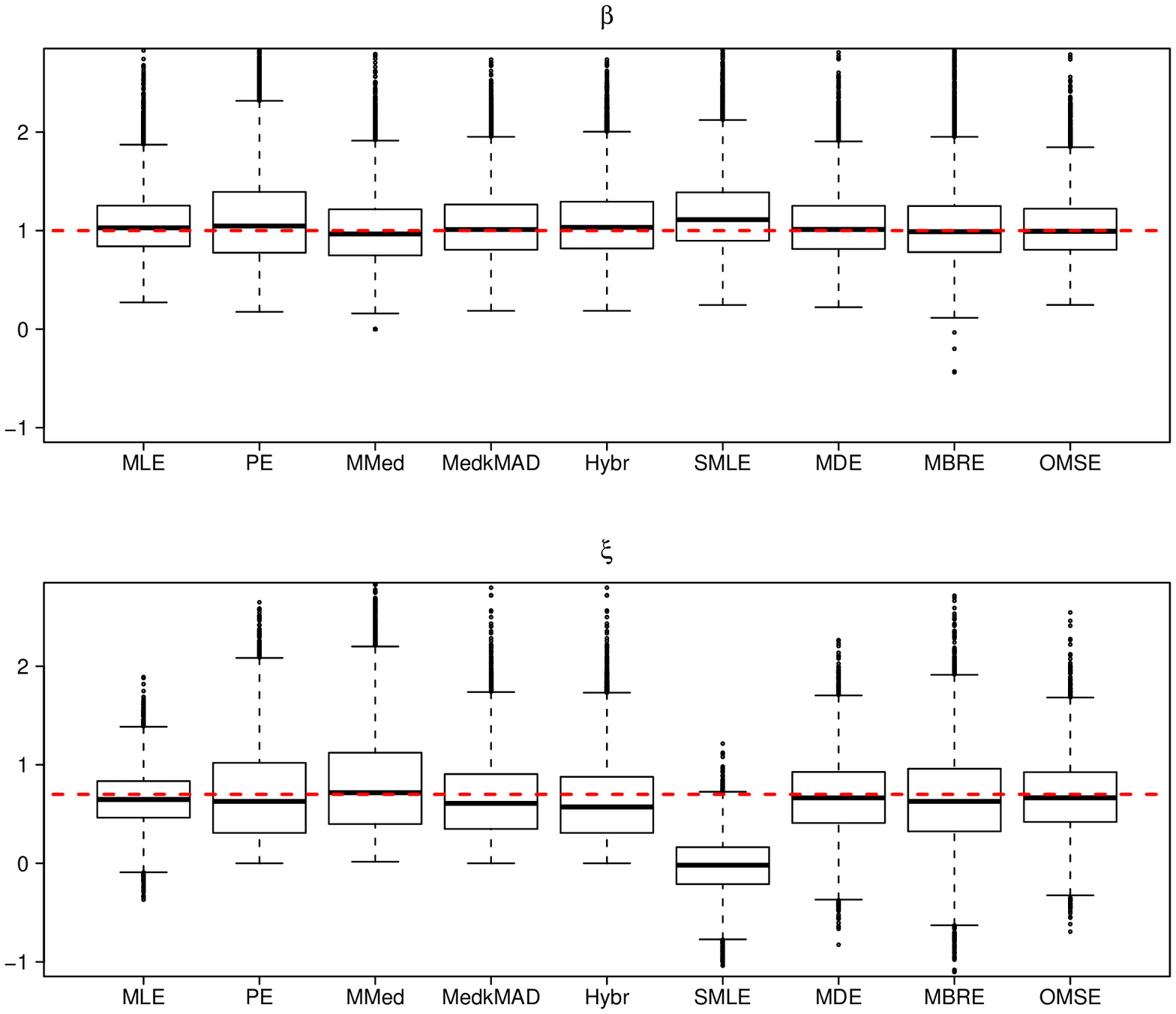}
\label{Fig:40}
}
\subfigure[7.9 \% contamination (corresponds to $r=0.5$), sample size $n=40$]{
\includegraphics[width=0.65\textwidth]{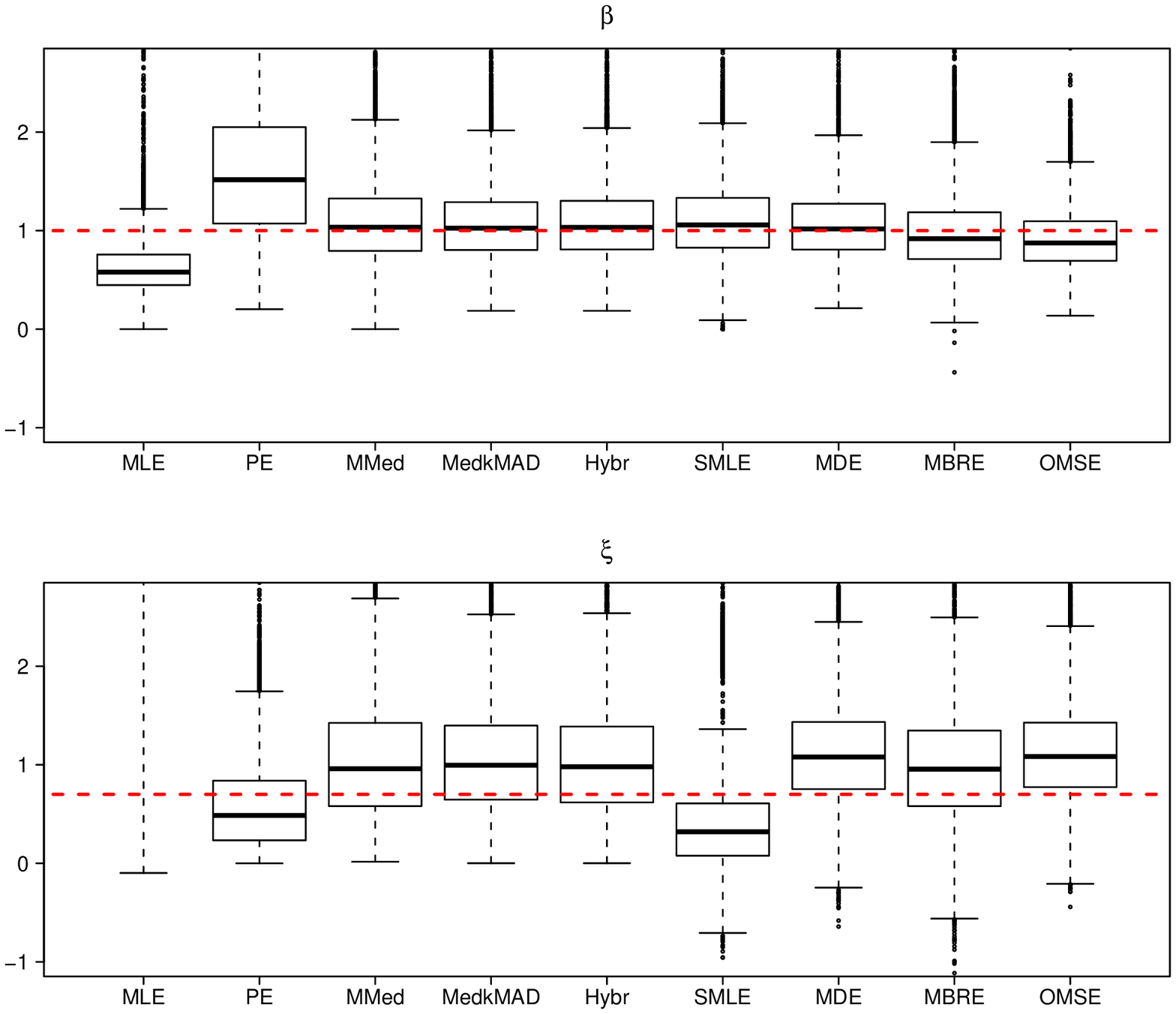}
\label{Fig:40_c}
}
\caption[Boxplots]{{Boxplots for
MLE, PE, MMed, MedkMAD, Hybr, SMLE (with $\approx 0.7\cdot \sqrt{40}$ skipped values),
MDE, MBRE, OMSE estimators for shape $\xi$ and scale $\beta$ of the GPD at ideal (above) and contaminated data (below),
\subref{Fig:40}, \subref{Fig:40_c};
number of runs: 10000; the red dashed line is the true parameter value.}}
\uPiii
\end{figure}

The results for sample size $40$ are illustrated in boxplots in
Figures~\ref{Fig:40} and \ref{Fig:40_c}, respectively. In Figure~\ref{Fig:40},
the underestimation of shape parameter $\xi$ by SMLE in the ideal situation
stands out; all other estimators in the ideal model are almost bias-free,
while PE is somewhat less precise; under contamination (Figure~\ref{Fig:40_c}),
all estimators are affected, producing bias, most prominently in coordinate
$\xi$. As expected, this effect is most pronounced for MLE which is completely
driven away, while the other estimators, at least in their medians stay near
the true parameter value.

\section{Application to Danish Insurance Data} \label{Appsec}
\begin{figure}[!htb]%[vt]
\uQiv
\centering
\includegraphics[width=8.5cm,height=0.9\textwidth,angle=270]{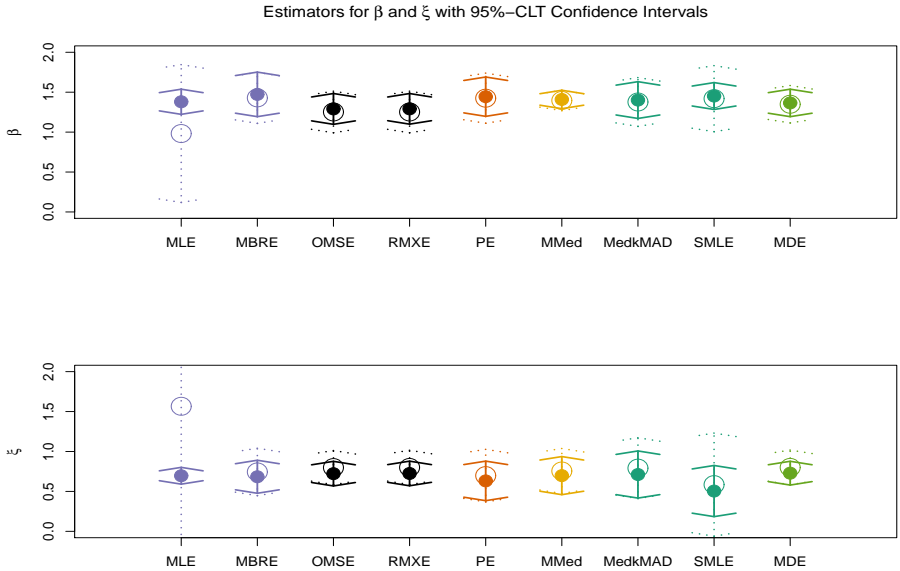}
\caption[Confidence plots]{\label{Fig:danish}{Confidence plots for
MLE, MBRE, OMSE, RMXE, PE, MMed, MedkMAD/Hybr, SMLE (with $\approx 0.7\cdot \sqrt{1000}$ skipped values),
MDE estimators for shape $\xi$ and scale $\beta$ of the GPD at ideal and contaminated data (solid/dashed arrows).
Confidence range for $\xi$ for MLE under contamination exceeds plotted region. \newline
Data: Danish insurance data set from {\sf R}-package {\tt evir} \citep{M:S:07}, threshold: $1.88\,{\rm MDKK}$, sample size $1000$,
contamination: $15$ data points modified to $10^{10}\,{\rm MDKK}$.
}}
\uPiv
\end{figure}

In Figure~\ref{Fig:danish} we illustrate the considered estimators evaluating them at the Danish
fire insurance data set from {\sf R} package~{\tt evir} \citep{M:S:07}.
This data set comprises $2167$ large fire insurance claims in Denmark from 1980 to 1990
collected at Copenhagen Reinsurance, supplied by  M.~Rytgaard of Copenhagen Re and
adjusted for inflation and expressed in millions of
Danish crowns (MDKK). For illustration purposes, we have chosen a threshold of
$1.88\,{\rm MDKK}$, leaving us $n=1000$ tail events. The values of estimates for shape and scale parameters
are plotted together with asymptotic $95\%$ (CLT-based) confidence intervals, denoted with
filled points and solid arrows respectively.
To visualize stability of the estimators against outliers at this data set, for
radius $r=0.5$, we artificially modify the original data set to a contaminated one
with $r \,\sqrt{n}$, or, after rounding, $15$ outliers with $10^{10}\,{\rm MDKK}$,
i.e.; an outlier rate of $1.5\%$.
The respective estimates on the contaminated data set are plotted with empty circles
and  confidence intervals with dashed arrows. For the contaminated data,
the confidence intervals are constructed to be bias-aware, i.e., with $\sqrt{\asMSE}$
instead of $\sqrt{\asVar}$ as scale. From Figure~\ref{Fig:danish} we can conclude,
that as expected, MLE is very sensitive to these $15$ outliers,
and that SMLE apparently tends to underestimate the shape parameter.
The OMSE, RMXE, and MDE produce reliable values not only for the original Danish data set,
but also for the contaminated one. MBRE and, worse, PE have a somewhat larger range of
variation, and MMed and MedkMAD (which coincides with Hybr here) for scale are quite well, but worse than the
OMSE, RMXE, and MDE for shape. Note that outliers at $10^{10}\,{\rm MDKK}$ are not
least favorable for PE and MMed.
% zur Kontrolle: as VaR on ideal data for (shape scale)
%           [,1]      [,2]
%  [1,]  2.876209  6.459820
%  [2,] 11.129523 20.265359
%  [3,]  6.189740  9.696972
%  [4,]  6.193580  9.579087
%  [5,] 15.962526 15.793984
%  [6,] 14.822576  3.466337
%  [7,] 22.684975 13.808726
%  [8,] 26.717321  7.378872
%  [9,]  5.775098  7.712579
% zur Kontrolle: as VaR on contaminated data for (shape scale)
%             [,1]      [,2]
%  [1,] 1340.16109 756.43410
%  [2,]   24.60576  26.96955
%  [3,]   12.17174  17.62875
%  [4,]   12.39185  17.69453
%  [5,]   27.51935  25.47028
%  [6,]   20.75629   4.41118
%  [7,]   36.21949  24.13481
%  [8,]  126.45797  41.43555

\section{Conclusion} \label{concl}
We have derived optimally robust estimators
MBRE, OMSE, and RMXE for scale and shape parameters $\xi$ and $\beta$ of the
GPD on ideal and contaminated data. Their computation has largely been accelerated
by interpolation techniques.

Among the potential starting estimators, clearly MedkMAD in its variant Hybr
excels and comes closest to the aforementioned group. %
%---${\rm eff.id} = {\rm eff.ru}= 40\%$, ${\rm eff.re}=78\%$.
%
For the same purpose, PE is also robust, but not really advisably due to its
low breakdown point and non-convincing efficiencies;
the only reason for using PE is its ease
of computation, which should not be so decisive.
Even worse is the popular SMLE without bias correction,
which does provide some, but much too little protection against
outliers.

Asymptotic theory and empirical simulations show that
Hybr, MedkMAD, MDE,  MBRE, OMSE, and RMXE  estimators
can withstand relatively high outlier rates as expressed by
an (E)FSBP of roughly $1/3$ (compare \citet{Ru:Ho:10,Ru:Ho:10b}).
SMLE in the variant without bias correction as used in this paper,
but with shrinking skipping rate, and MLE have minimal
FSBP of $1/n$, hence should be avoided.

High failure rates for MMed and MedkMAD for small $n$,
and under contamination limit their usability considerably,
while Hybr works reliably.

Looking at the influence functions, we see that,
except for MLE,  all estimators have bounded $\IF$s, so finite
GES, but %. As visible in Figure~\ref{IC-fig}, the estimators
do differ in how they use the information contained in an observation.

This is reflected in asymptotic values, as well as in (simulated)
finite sample values: for known radius we can recommend  OMSE with
{\rm Hybr} as initialization. It has best statistical properties in
the simulations, is computationally fast, efficient for contamination
of known radius. % and,  for varying $\xi$, never drops below
%$58\%$ efficiency in the ideal model and for contamination of unknown radius
%(see Table~\ref{tabmin}).
%
 MBRE, and MDE come close to OMSE. % with efficiencies
% ${\rm eff.id} = {\rm eff.ru}= 41\%$, ${\rm eff.re}=78\%$ (MBRE),
%and ${\rm eff.id} = 45\%$, ${\rm eff.re}=69\%$, ${\rm eff.ru}=43\%$ (MDE).
%
For unknown radius RMXE % with ${\rm eff.id} = {\rm eff.ru}= 63\%$, ${\rm eff.re}=98\%$
is recommendable with again OMSE, MBRE,  Hybr and MDE (in this order) as close
competitors.

All estimators are publicly available in {\sf R} on \texttt{CRAN}.

%
%The worst as to all robustness aspects is MLE.

%\section*{Acknowledgement}
%The authors thank an anonymous referee of a previous revision
%of this article for his helpful and valuable comments.

\appendix
\small
%----------------------------------------------------------------------------------
\section{Estimators}\label{EstimatorsApp}
%----------------------------------------------------------------------------------
For each of the estimators discussed in Section~\ref{Estimators}, we
determine its IF, its asymptotic variance {\rm asVar}, its maximal asymptotic bias {\rm asBias},
and its FSBP where possible. All estimators considered in this appendix
are defined in the original ($\beta$-)scale and equivariant in the sense of \eqref{scaleeq}.

%----------------------------------------------------------------------------------
\subsection{Estimators Obtained as Minima or Maxima}%\vspace{-1ex}
%----------------------------------------------------------------------------------
\begin{proposition}[\textbf{(MLE)}]\label{MLEprop}
\begin{enumerate}
\item[{\textrm{\bf IF}}] $\IF_\theta(x;{\rm MLE},F_\theta) =  \mathcal{I}_\theta^{-1} \Lambda_\theta(x)$.
where, using the quantile-type representation~\eqref{vdef}
\begin{equation} \label{tpsidef}
\tilde\psi(v) =
\Tfrac{\xi+1}{\xi^2}\Big(\begin{array}{c} -(\xi^2+\xi)\log(v)+
(2\xi^2+3\xi+1)v^\xi-(\xi^2+3\xi+1)\\
\xi\log(v)-(2\xi^2+3\xi+1)v^\xi+(3\xi+1)
\end{array}\Big)
\end{equation}
MLE attains the smallest asymptotic variance among all ALEs.
\item[{\textrm{\bf asVar}}]
\begin{equation}
\asVar({\rm MLE})={\mathcal{I}_\theta}^{-1} =
(1+\xi)\left(\begin{array}{cc}
 \xi+1,& -\beta \\
-\beta, & 2\beta^2
\end{array}\right)
\end{equation}
\item[{\textrm{\bf asBias}}]
Both components of the joint IF are unbounded---although only
growing in absolute value at rate $\log(x)$.
\item[{\textrm{\bf FSBP}}]
The FSBP of MLE is minimal, i.e.; $1/n$.
\end{enumerate}
\end{proposition}
%
%--------------------------------------------------------------------------------
%\subsection{Skipped Maximum Likelihood Estimators}\label{SMLEsec}%\vspace{-1ex}
%--------------------------------------------------------------------------------
%
As we have seen, SMLE in fact does not estimate $\theta$ but
$d(\theta)=\theta + B_\theta$, for bias $B_\theta$ already present in the ideal model.
\begin{proposition}[\textbf{(SMLE)}]\label{SMLEprop}
\begin{enumerate}
\item[{\textrm{\bf IF}}]
The functional $T(F_\theta):=SMLE(F_\theta)+B_\theta$ estimating $d(\theta)$ may be written as
\begin{align}
T(F)&= \frac{1}{1-\alpha}\int_{0}^{1-\alpha} \Lambda_\theta(F^{-1}(s))ds \label{SMLEfunc}
\end{align}
With $u_\alpha:=F^{-1}(1-\alpha)$, its IF is given by
\begin{align}
\IF_\theta(z;{\rm  T},F_\theta) &= {\mathcal{I}_\theta}^{-1} \left\lbrace
\begin{array}{ll}
\frac{1}{1-\alpha}[\Lambda_\theta(z)- W(F)], &  0\leq x \leq u_\alpha\\
\frac{1}{1-\alpha}[\Lambda_\theta(u_\alpha)- W(F)],  &x > u_\alpha\\
\end{array} \right. \label{SMLE1}\\
W(F) &= (1-\alpha)\mathop{\rm T}(F) +  \alpha \Lambda_\theta(u_\alpha) \label{SMLE2}
\end{align}
\item[{\textrm{\bf asVar}}] Numeric values can be obtained by integrating out  $\IF_\theta(z;{\rm  T},F_\theta)$.
\item[{\textrm{\bf asBias}}]
For shrinking rate $\alpha_n=r'/\sqrt{n}$,
asymptotic bias of SMLE is finite for each $n$, but, standardized by $\sqrt{n}$,
is of exact order $\log(n)$, hence unbounded. The bias induced by contamination
is dominated by $B_{n,\theta}$ eventually in $n$.
\item[{\textrm{\bf FSBP}}]
FSBP$=\alpha_n$ eventually in $n$.
\end{enumerate}
\end{proposition}
%--------------------------------------------------------------------------------
%\subsection{Cram\'er-von-Mises Minimum Distance Estimators}%\vspace{-1ex}
%--------------------------------------------------------------------------------

\begin{proposition}[\textbf{(MDE)}]\label{MDEprop}
\begin{enumerate}
\item[{\textrm{\bf IF}}]
For $v$ from \eqref{vdef}, the IF of MDE is given by
\begin{eqnarray}
\IF(x;{\rm MDE},F_\theta) &=& 3(\xi+3)^2\left(\begin{array}{cc}
\Tfrac{18(\xi+3)}{(2\xi+9)}, & -3\beta \\
-3\beta, & 2\beta^2
\end{array}\right)\left(\begin{array}{c}
\tilde \varphi_\xi\\
\tilde \varphi_\beta
  \end{array}\right)(v(z(x))),\qquad \mbox{for}\nonumber\\
\left(\begin{array}{c}
\tilde \varphi_\xi\\
\tilde \varphi_\beta
  \end{array}\right)(v)&=&
\left(\begin{array}{l}
\Tfrac{19+5\xi}{36(3+\xi)(2+\xi)}
+\Tfrac{1}{\xi} v^{2} \log(v) + \Tfrac{2-\xi}{4\xi^2} v^{2}
- \Tfrac{1}{\xi^2(2+\xi)} v^{2+\xi}\\
\Tfrac{5+\xi}{6(3+\xi)(2+\xi)\beta}
- \Tfrac{1}{2\xi\beta} v^{2}
+ \Tfrac{1}{\xi\beta(2+\xi)}v^{2+\xi}
\end{array}\right)
\end{eqnarray}
\item[{\textrm{\bf asVar}}]
\begin{align}
\asVar({\rm MDE})&=\frac{\left( 3+\xi \right) ^{2}}{125\left( 5+2\,\xi \right)
\left( 5+\xi \right) ^{2} }\left(\begin{array}{cc}
  V_{1,1},&V_{1,2}\\
  V_{1,2},&V_{2,2}
  \end{array}\right)\qquad
%\end{equation}
%
%\noalign
\mbox{for} \\
\nquad \Ts V_{1,1}&= {81\left(
16 {\xi}^{5}+272 {\xi}^{4}+ 1694 {\xi}^{3}+4853 {\xi}^{2}+7276 \xi+6245
 \right)
 (2 \xi+9 )^{-2}},
\nonumber\\
\nquad \Ts  V_{1,2}&=- {9 \beta
\left( 4 {\xi}^{4}+86 {\xi}^{3}+648 {
\xi}^{2}+2623 \xi+4535 \right)
 ( 2 \xi+9 )^{-1}   },
\nonumber\\
\nquad \Ts V_{2,2}&={\beta}^{2}
 \left( 26 {\xi}^{3}+601 {\xi}^{2}+3154 \xi+5255 \right)\nonumber
\end{align}
\item[{\textrm{\bf asBias}}]
$\asBias({\rm MDE})$ is finite.
\item[{\textrm{\bf FSBP}}]
The FSBP of MDE is at least $1/2$ of the optimal FSBP achievable in this
context. An upper bound is given by
\begin{equation}\label{BDPbound}
\ve^\ast_n \leq \min \Big\{ \,\Tfrac{-\inf_{v,\xi}  %
\tilde \varphi_{\scriptscriptstyle\Bullet}}%
{\sup_{v,\xi} \tilde\varphi_{\scriptscriptstyle\Bullet}-%
\inf_{v,\xi}  \tilde\varphi_{\scriptscriptstyle\Bullet}},\quad %\\
\Tfrac{\sup_{v,\xi} \tilde\varphi_{\scriptscriptstyle\Bullet}}%
{\sup_{v,\xi} \tilde\varphi_{\scriptscriptstyle\Bullet}-%
\inf_{v,\xi}  \tilde\varphi_{\scriptscriptstyle\Bullet}},
\quad \Bullet=\xi,\beta \,\Big\}
\end{equation}
\end{enumerate}
\end{proposition}

\begin{figure}[!htb]
\uQv
\centering
\includegraphics[bb=18 18 594 774,height=0.26\textheight,width=0.52\textwidth]{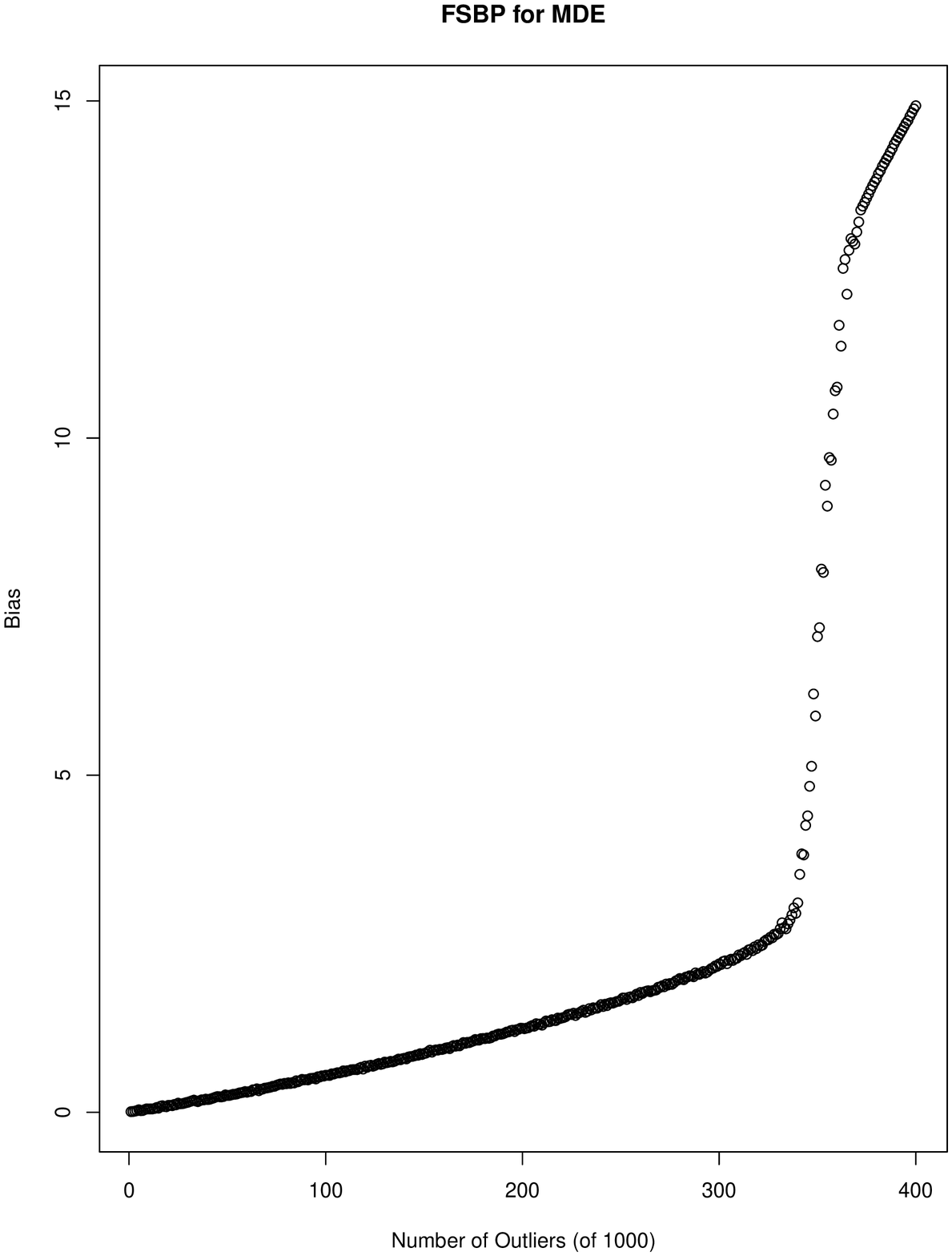}
{\flushleft\caption{\label{Fig:MDE_FSBP}{Empirical Bias for FSBP of MDE to CvM distance\newline
This bias is computed by simulating $M=100$ samples of size $n=1000$ from a GPD with
 $\xi=0.7$, $\beta=1$, and after replacing $m$ observations, for
 $m=1,\ldots, 400$ by value $10^{10}$.
%compute the bias.
There is  a steep increase around  $354$, so we conjecture that (E)FSBP should be
approximately $0.35$.}}}
\uPv
\end{figure}

To make the inequality in \eqref{BDPbound}
an equality, we would need to show that we cannot produce a
breakdown with less than this bound.
Evaluating bound~\eqref{BDPbound} numerically gives a value of
$4/9\doteq 36\%$, which is achieved
for $v=0$ (and $\xi\to 0$) or, equivalently, letting the $m$
replacing observations in Definition~\eqref{FSBPDef}
tend to infinity.
%
%\begin{Rem}
To see how realistic this value is compare Figure \ref{Fig:MDE_FSBP},
where we produce an empirical max-bias-curve by simulations.
%, simulating
%$M=100$ samples of size $n=1000$ observations from a GPD with
%$\xi=0.7$, $\beta=1$, and after replacing $m$ observations, for
%$m=1,\ldots, 400$  by value $10^{10}$
%compute the bias.  There is  a steep
%increase around  $354$, so we conjecture that (E)FSBP should be
%approximately $0.35$; on the other side, MDE cannot have a higher
%FSBP than its initialization.
%, and so far the best known initialization has (E)FSBP of $0.346$.
%\end{Rem}

%
%----------------------------------------------------------------------------------
\subsection{Starting Estimators}%\vspace{-1ex}
%----------------------------------------------------------------------------------
%--------------------------------------------------------------------------------
%\subsection{Pickands Estimator} \label{PEsec}\vspace{-1ex}
%--------------------------------------------------------------------------------
%
\begin{proposition}[\textbf{(PE)}]\label{PEprop}
\begin{enumerate}
\item[{\textrm{\bf IF}}]
\begin{align}
\IF_\Bullet (x;{\rm PE}(a),F_\theta) &=   \sum\nolimits_{i = 2,3} h_{\Bullet,i}(a)
\Tfrac{\alpha_i(a) - \mathbb{I}(x \leq \hat Q_{i}(a))}{f(\hat Q_{i}(a))}, \qquad \Bullet = \xi,\beta %\\
\end{align}
with deterministic (signed) weights $h_{\Bullet,i}(a)$ to given in the proof.
\item[{\textrm{\bf asVar}}]
Abbreviating $\alpha_i(a)$ by $\alpha_i$, $1-\alpha_i$ by $\bar \alpha_i$,
and $h_{\Bullet,i}(a)$ by $h_{\Bullet,i}$, %$\Bullet = \xi,\beta$,
the asymptotic covariance for PE(a) is
\begin{align}
&\asVar({\rm PE}(a)) = \beta^2 \left(\begin{array}{cc}
h_{\xi,2}, & h_{\beta,2} \\
h_{\xi,3}, & h_{\beta,3}
\end{array}\right)
 \left(\begin{array}{cc}
\alpha_2 {\bar\alpha_2}^{-1-2\xi},&
\alpha_2 {\bar\alpha_2}^{-1-\xi}{\bar\alpha_3}^{-\xi}\\
\alpha_2 {\bar\alpha_2}^{-1-\xi}{\bar\alpha_3}^{-\xi},&
\alpha_3 {\bar\alpha_3}^{-1-2\xi}
\end{array}
\right) \left(\begin{array}{cc}
h_{\xi,2}, & h_{\xi,3} \\
h_{\beta,2}, & h_{\beta,3}
\end{array}\right)
\end{align}
\item[{\textrm{\bf asBias}}]
$\asBias({\rm PE})$ is finite.
\item[{\textrm{\bf FSBP}}]
$ %\begin{equation} \label{Nn0def}
\ve_n^\ast =\min\{1/a^2,\hat N^0/n\}$, for
$\hat N^0_n:=\#\{X_i\,\big|\,2 \hat Q_2(a)\leq X_i\leq \hat Q_3(a)\}
$.\\  %\end{equation}
%
%and $\hat N^0/n\to \pi_\xi(a)=(2a^\xi-1)^{-1/\xi}-1/a^2$, so that
%
$%\begin{equation}
\bar \ve^\ast = \bar \ve^\ast(a)= \min\{\pi_\xi(a),1/a^2\}
$ %\end{equation}
for $\pi_\xi(a)=(2a^\xi-1)^{-1/\xi}-1/a^2$.
\end{enumerate}
\end{proposition}
For $\xi=0.7$,  the classical PE achieves an  ABP of
$\bar \ve^\ast(a=2)\doteq 6.42\%$; as to EFSBP, for  $n=40,100,1000$
we obtain $\bar \ve^\ast_n= 5.26\%,6.34\%,6.42\%$, respectively
\citep[Table~2]{Ru:Ho:10b}.
%--------------------------------------------------------------------------------
%\subsection{Method of Medians Estimator} \label{MMedsec}\vspace{-1ex}
%--------------------------------------------------------------------------------
%\pagebreak
\begin{proposition}[\textbf{(MMed)}]\label{MMedprop}
\begin{enumerate}
\item[{\textrm{\bf IF}}] Let $M(\xi):=\mbox{\rm $\Lambda$-Med}(F_{\theta_1})={\rm median}(\Lambda_{\theta_1;2}\circ F_{\theta_1})$
the population median of the shape scores, $l_i:=\frac{\partial}{\partial x} \Lambda_{\theta_1;2}(q_i)$, and
$m=m_\xi:=F^{-1}_{\theta_1}$ the population median.
Then the level set $\{x\in\R\,\mid\,\Lambda_{\theta_1;2}(x)\leq M(\xi)\}$
is of form $[q_1(\xi),q_2(\xi)]$ and
$% \begin{equation}
{\rm IF}(x;{\rm MMed},F_\theta)=D({\rm IF}(x;{\rm median},F_\theta),%
                           {\rm IF}(x;\mbox{\rm $\Lambda$-Med},F_\theta))^\tau
$ %\end{equation}
where
\begin{equation}
\IF(x;{\rm median},F_\theta) = \Big(\Tfrac{1}{2} - \mathbb{I}(x\leq m)\Big)/f(m),\qquad
\IF(x;\mbox{\rm $\Lambda$-Med},F_\theta)=\frac{\mathbb{I}(q_1\leq x\leq q_2)-1/2}%
{f_\theta(q_2)/l_2-f_\theta(q_1)/l_1} \label{lmmedA}
\end{equation}
and $D$ is a corresponding deterministic Jacobian.
\item[{\textrm{\bf asVar}}]
Let
\begin{equation} \label{tildeDdef}
\tilde D^{-1}= \Ew_\theta \chi_\theta\Lambda_\theta^\tau \ \mbox{ for } \
\chi_\theta(x)=d_\beta \chi_{\theta_1}(\Tfrac{x}{\beta}),\quad
\chi_{\theta_1}(x)=\Big(\mathbb{I}(x\leq m_\xi)-1/2,\;%
                    \mathbb{I}(q_1\leq x\leq q_2)-1/2\Big)^\tau
\end{equation}
Then
\begin{align}
\asVar({\rm MMed}) &= \frac{1}{4}\,\tilde D
\left(\begin{array}{cc}
1,&
1-4F(q_1)\\
1-4F(q_1),&
1
\end{array}
\right) \tilde D^\tau
\end{align}
\item[{\textrm{\bf asBias}}]
$\asBias({\rm MMEd})$ is finite.
\end{enumerate}
\end{proposition}
We have not found analytic breakdown point values, neither for ABP nor for FSBP.
While $50\%$ by scale equivariance is an upper bound, the high frequency of failures in the simulation
study for small sample sizes however indicates that (E)FSBP should be
considerably smaller; a similar study for the empirical maxBias as the one
for MDE  gives that for sample size $n$ from a rate of outliers of
$\epsilon_n$ on, we have but failures in solving for MMed,
for $\epsilon_{40}=42.5\%$, $\epsilon_{100}=35.0\%$,
$\epsilon_{1000}=25.1\%$, and $\epsilon_{10000}=20.1\%$.
So we conjecture that the asymptotic breakdown point $\ve^\ast\leq 20\%$.

%
%--------------------------------------------------------------------------------
%\subsection{MedkMAD}\label{Sec:MedkMAD}\vspace{-1ex}
%--------------------------------------------------------------------------------

\begin{proposition}[\textbf{(MedkMAD)}]\label{MedkMADprop}
\begin{enumerate}
\item[{\textrm{\bf IF}}] Let $G=G\left( (\xi,\beta);(M,m) \right)$ be the
defining equations of MedkMAD, i.e.;
\begin{equation}
G\left( (\xi,\beta);(M,m) \right) = (G^{(1)},  G^{(2)})^\tau =
\Big(f_{m,\xi,\beta;k}(M),\;
\beta\Tfrac{2^\xi-1}{\xi} - m\Big)^\tau
\end{equation}
and
\begin{equation} \label{DdefMedkMAD}
D= -\Big(\Tfrac{\partial G}{\partial (\xi,\beta)}\Big)^{-1}
\Tfrac{\partial G}{\partial (M,m)}
\end{equation}
Then the IF of MedkMAD estimator is
$%\begin{align}
\IF(x;{\rm MedkMAD},F_\theta) = D\,  (\IF(x;{\rm kMAD},F_\theta),\IF(x;{\rm median},F_\theta))^\tau
$ %\end{align}
where the IF of kMAD is given by
\begin{equation}
\IF(x;{\rm kMAD},F_\theta) =
\Tfrac{\frac{1}{2} -
\mathbb{I}(-M\leq x-m\leq kM)}{f(m + kM) - f(m - M)}+
\Tfrac{f(m + kM) - f(m - M)}{kf(m + kM) + f(m - M)}
\Tfrac{\mathbb{I}(x\leq m) - \frac{1}{2}}{f(m)} \label{cpdef}
\end{equation}

\item[{\textrm{\bf asVar}}] Let $a_s := f(m-M) +s f(m+kM)$, and $d = a_{-1}^2 + 4(1-a_1) a_{-1} f(m)$
and
\begin{equation}
\sigma_{1,1} = (4f(m))^{-2}, \quad \sigma_{2,2} =
\Tfrac{f(m)^2}{4 a_k^2 (f(m)^2 + d)},\quad
\sigma_{1,2} = \sigma_{2,1} = \Tfrac{1 - 4 F(m-M) +  a_{-1}/f(m)}{4 f(m) a_k}
\end{equation}
Then
\begin{equation}
\asVar({\rm MedkMAD})=D^\tau
\left(\begin{array}{cc}
\sigma_{1,1},& \sigma_{1,2}\\\sigma_{2,1}&\sigma_{2,2}
\end{array}\right)
D
\end{equation}
\item[{\textrm{\bf asBias}}]
$\asBias({\rm MedkMAD})$ is finite.
\item[{\textrm{\bf FSBP}}]
%\begin{equation}
$
\ve_n^\ast =\min\{\hat N'_n,\hat N''_n\}/n
$
 %\end{equation}
\mbox{ for}
%\begin{eqnarray}
\begin{equation}
\hat N_n'=\#\{X_i \,| \hat m < X_i\leq (k+1)\hat m\,\},\qquad
\hat N_n''=\lceil n/2\rceil - \#\{X_i \,|\, (1-\check q_k) \hat m \leq %
X_i \leq (k\check q_k+1)\hat m \}\label{Nn2def}
\end{equation}
%\end{eqnarray}
and
\begin{equation}
\bar \ve^\ast = \min\Big(F_\theta((k\!\!+\!\!1)m)-
\Tfrac{1}{2}, \;F_\theta\big((k\check q_k\!+\!1)m\big)-
F_\theta\big((1\!-\!\check q_k)m\big)-\Tfrac{1}{2}\Big)
\end{equation}
\end{enumerate}
\end{proposition}

For $\xi=0.7$, the EFSBP is given by the first alternative if $k<3.23$ and
by the second one otherwise.

As to the choice of $k$, it turns out that a value of $k=10$ gives
reasonable values of ABP, asVar, asBias for a wide range of parameters
$\xi$, see \citet{Ru:Ho:10}. %, as documented in Table~\ref{MedkMADTab}.
In the sequel this will
be our reference value for $k$; as to EFSBP,
for $n=40,100,1000$ and $\xi \in \mathbb{R}$ we obtain $\bar \ve^\ast_n= 42.53\%, 43.86\% , 44.75\%$,
respectively \citep[Table~2]{Ru:Ho:10b}.
%
%\begin{table}[ht]
%\begin{center}
%\begin{tabular}{c|rr|rr|rr|rr}
%  \hline
% $\xi$&  ${\rm GES}$ &  ${\rm GES}^{\ssr opt}$ & $\asVar$ & $\asVar^{\ssr opt}$ & %
% $\asMSE$ & $\asMSE^{\ssr opt}$ & ${\rm ABP}$ &${\rm ABP}^{\ssr opt}$\\
%  \hline
%$0.01$ & $ 4.09$ & $ 2.71$ & $12.08$ & $ 3.04$ & $16.26$ & $ 7.58$ & $0.249$& $0.322$\\
%$0.10$ & $ 3.83$ & $ 2.84$ & $10.90$ & $ 3.41$ & $14.58$ & $ 8.39$ & $0.259$& $0.325$\\
%$0.70$ & $ 4.38$ & $ 3.66$ & $12.80$ & $ 6.29$ & $17.60$ & $14.13$ & $0.310$& $0.342$\\
%$1.50$ & $ 5.85$ & $ 4.82$ & $19.50$ & $11.25$ & $28.06$ & $24.03$ & $0.355$& $0.358$\\
%$4.00$ & $10.58$ & $ 8.42$ & $52.90$ & $35.00$ & $80.90$ & $56.86$ & $0.221$& $0.379$\\
%   \hline
%\end{tabular}
%\caption{\label{MedkMADTab} {Robustness properties of MedkMAD for $k=10$ and
%several shape parameters compared to corresponding optimal values, i.e.; %
%MBRE (${\rm GES}$), MLE ($\asVar$), OMSE ($\asMSE$), ${\rm MedkMAD}(k^{\ssr ABP})$, %
%$k^{\ssr ABP}=\mathop{\rm argmax}_k {\rm ABP}({\rm MedkMAD}(k))$ (ABP)}}
%\end{center}
%\uP
%\end{table}
%
Results on optimizing MedkMAD in $k$ w.r.t.\ the different robustness criteria
for $\xi=0.7$ can be looked up in  \citet[Table~5]{Ru:Ho:10}.

% \begin{Rem}
%  Admittedly,  for given $k$, eventually in $n$,
% $\Ew_{(\xi,\beta)}[\ve_n^\ast({\rm MedkMAD}(k))]$
% is decreasing in $\xi$ s.t.\
% $
% \lim_{\xi\to\infty} \Ew_{(\xi,\beta)}[\ve_n^\ast({\rm MedkMAD}(k))]=0
% $.
% At the same time, eventually in $n$,
% $\xi\mapsto  \Ew_{(\xi,\beta)}[\ve_n^\ast({\rm PE*})]$ is increasing with
% $
% \lim_{\xi\to\infty} \Ew_{(\xi,\beta)}[\ve_n^\ast({\rm PE*})]=1/4
% $.
%  In particular, for $k=10$, for $\xi\ge 4.964$,
% PE* has a better EFSBP / ABP, in this case $\bar\ve^\ast({\rm PE*})\ge19.0\%$.
% But,  eventually in $n$, the EFSBP of MedkMAD
% for the ABP-optimal $k^{\ssr ABP}=k^{\ssr ABP}(\xi)$ never drops below
%  $32.1\%$ for $\xi \in(0,10]$ and
% below $25\%$ for $\xi\in(0,437]$, and achieves $39.9\%$ for $\xi=7.20$.
% \end{Rem}
%
\section{Proofs}\label{proofsec}
To assess integrals in the GPD model the following lemma is helpful,
the proof of which follows easily by noting that $v(z)$
introduced in it is just the quantile
transformation of ${\rm GPD}(0,\xi,1)$ up to the flip $v\mapsto 1-v$.

\begin{lemma}\label{Hilfslem}
Let $X\sim {\rm GPD}(\mu,\xi,\beta)$ and let $z=z(x)=(x-\mu)/\beta$ and
\begin{equation} \label{vdef}
v=v(z)=(1+\xi z)^{-1/\xi}
\end{equation}
Then for $U\sim {\rm unif}(0,1)$, we obtain ${\cal L}(v(U))={\rm GPD}(0,\xi,1)$
and ${\cal L}(\beta v(U) + \mu)={\cal L}(X)$.
\end{lemma}

\mparagraph{Proof of Proposition~\ref{smoothmodel}: }
%\begin{proof}
We start by differentiating the log-densities $f_\theta$ pointwise in $x$
w.r.t.\ $\xi$ and $\beta$ to obtain \eqref{LBdef} and, using
Lemma~\ref{Hilfslem} we obtain
the expressions for \eqref{FIdef}, from where we see finiteness and positive definiteness.
As density $f_\theta$ is differentiable
in $\theta$ and the corresponding Fisher information is finite and
continuous in $\theta$, by \citet[App.~A]{Ha:72}, %\citet[Satz~1.194]{Wit:85}
this entails $L_2$-differentiability.
\hfill
            \QEDlogo
            %\qed
%\end{proof}
\smallskip

\mparagraph{Proof of Lemma~\ref{InvProp}: }
For the first half of \eqref{InvProp1} let $h=(h_\xi,h_\beta)$ and $h'=d_\beta^{-1} h$.
We note that $f_\theta(x)=f_{\theta_1}(x/\beta)/\beta$,
hence $f_{\theta+h}(x)=f_{\theta_1+h'}(x/\beta)/\beta$.
Then
\begin{eqnarray*}
&&\int \left(f^{1/2}_{\theta+h}(x)-f^{1/2}_{\theta}(x)(1+\Tfrac{1}{2} \Lambda^{\tau}_{\theta_1}(\Tfrac{x}{\beta})d_\beta^{-1}h)\right)^2\,dx=\\
&=&
\int \Tfrac{1}{\beta}\left(f^{1/2}_{\theta_1+h'}(\Tfrac{x}{\beta})-f^{1/2}_{\theta_1}(\Tfrac{x}{\beta})(1+\Tfrac{1}{2} \Lambda^{\tau}_{\theta_1}(\Tfrac{x}{\beta})h')\right)^2\,dx=\\
&=&\int \left(f^{1/2}_{\theta_1+h'}(y)-f^{1/2}_{\theta_1}(y)(1+\Tfrac{1}{2} \Lambda^{\tau}_{\theta_1}(y)h')\right)^2\,dy=\Lo(|h'|^2)=\Lo(|h|^2)
\end{eqnarray*}
So indeed, the $L_2$-derivative $\Lambda_{\theta}(x)$ is given by $d_\beta^{-1}\Lambda_{\theta_1}(x/\beta)$.
Equation~\eqref{Lbtransf} is a consequence of the chain rule. This also entails the second half
of \eqref{InvProp1}:
$
\tilde\Lambda_{\tilde\theta}(x)= d_\beta\Lambda_{\theta}(x)=
\Lambda_{\theta_1}(x/\beta) =  \tilde\Lambda_{\tilde\theta_0}(x/\beta)
$.
The assertions for ${\cal I}_\theta$, $\tilde {\cal I}_{\tilde \theta}$ are simple consequences.
\hfill
            \QEDlogo
\pagebreak

\mparagraph{Proof of Proposition~\ref{scalinvprop}: }
\begin{itemize}
\item[(a)] Paralleling the proofs to \citet[Thm.'s 5.5.7, 5.5.1, and Lem.~5.5.10]{Ried:94}, we see that
 the assertions of the theorems are also valid for general norms derived from quadratic
forms; the only place leading to visible modification of the result is determining clipping height $b$ of $\bar\psi$.
In the proof of Thm.~5.5.1, the expression corresponding to $\tr A$ arises
as $\Ew \bar \psi^\tau d_\beta^{-2} Y=\tr d_\beta^{-2} \Ew Y \psi^\tau =\tr d_\beta^{-2} A$.
\item[(b)] With the definitions of $A_\theta$, $a_\theta$, $b_\theta$ from \eqref{Aeqvar}, we obtain
\begin{eqnarray*}
Y_{\theta}(x)&=&A_{\theta} \Lambda_{\theta}(x) - a_{\theta}=
d_\beta A_{\theta_1} d_\beta d_\beta^{-1} \Lambda_{\theta_1}(\Tfrac{x}{\beta})- d_\beta a_{\theta_1}=%\\
%&=&
d_\beta Y_{\theta_1} (\Tfrac{x}{\beta})
\end{eqnarray*}
so in particular $n_\beta\big(Y_{\theta}(x)\big)
=n_1\big(Y_{\theta_1}(x/{\beta})\big)$.
For \eqref{bet=1lsg}, we hence only have to check that, starting with the optimal IF $\psi_{\theta_1}\in \Psi_2(\theta_1)$, function
$\psi^{(0)}(x):=d_\beta \psi_{\theta_1}(x/\beta)\in \Psi_2(\theta)$ and
solves \eqref{barpsidef} respectively \eqref{hatpsidef}. By Lemma~\ref{InvProp} and with $X'\sim{\rm GPD}(\theta_1)$ and $X=\beta X'$, we get
\begin{eqnarray*}
&&\Ew_{\theta}\psi^{(0)}(X)=d_\beta \Ew_{\theta} \psi_{\theta_1}(\Tfrac{X}{\beta})=d_\beta \Ew_{\theta_1} \psi_{\theta_1}(X')=0\\
&&\Ew_{\theta}\psi^{(0)}(X)\Lambda^{\tau}_{\theta}(X)=
d_\beta \Ew_{\theta} \psi_{\theta_1}(\Tfrac{X}{\beta})\Lambda^{\tau}_{\theta_1}(\Tfrac{X}{\beta})d_\beta^{-1}=
d_\beta \Ew_{\theta_1} \psi_{\theta_1}(X')\Lambda^\tau_{\theta_1}(X')d_\beta^{-1}=\EM_2
\end{eqnarray*}
To see that $b_{\theta}=b_{\theta_1}$, for \eqref{barpsidef} we see that with $A'=d_\beta A d_\beta$ and $a' = d_\beta a$
\begin{eqnarray*}
b_{\theta} &=& \max_{A,a} \frac{\tr d_\beta^{-2} A }{ \Ew_{\theta} n_\beta\Big(A\Lambda_{\theta}(X)-a\Big)}=
\max_{A',a'} \frac{\tr A'}{ \Ew_{\theta} n_1\Big(A'\Lambda_{\theta_1}(\Tfrac{X}{\beta})-a'\Big)}=\\
&=&
\max_{A',a'} \frac{\tr A' }{ \Ew_{\theta_1} n_1\Big(A'\Lambda_{\theta_1}(X')-a'\Big)}= b_{\theta_1}
\end{eqnarray*}
while for \eqref{hatpsidef} this follows from
\begin{eqnarray*}
r^2 b_{\theta_1} &=& \Ew_{\theta_1} \Big(n_1\big(Y_{\theta_1}(X')\big)-b_{\theta_1}\Big)_+=\Ew_{\theta} \Big(n_1\big(Y_{\theta_1}(\Tfrac{X}{\beta})\big)-b_{\theta_1}\Big)_+=
\Ew_{\theta} \Big(n_\beta \big(Y_{\theta}(X)\big)-b_{\theta_1}\Big)_+
\end{eqnarray*}
\item[(c)] Similarly as in (b), denoting by $\tilde\Psi_2$ the set of IFs in the log-transformed model,
we have to check that starting from the optimal IF $\eta_{\tilde\theta_0}\in\tilde\Psi_2(\theta_0)$ function
$\eta^{(0)}(x):=\eta_{\tilde\theta_0}(x/\beta)\in \tilde\Psi_2(\tilde \theta)$ and
solves \eqref{barpsidef} respectively \eqref{hatpsidef}; but by Lemma~\ref{InvProp}, this follows by analogue
arguments as in (b).
\item[(d)] Again we have to show that for optimally-robust IF $\psi_{\theta}\in\Psi(\theta)$
function $\eta^{(0)}:=d_{\beta}^{-1}\psi_{\theta}\in \tilde \Psi(\tilde\theta)$ and
solves \eqref{barpsidef} respectively \eqref{hatpsidef} in the log-scale model; but by \eqref{Lbtransf},
this is shown like in (b).
\hfill       \QEDlogo%\end{proof}
\end{itemize}

\smallskip

\mparagraph{Proof of Lemma~\ref{Lemma_exp_scale}: }
%\begin{proof}
Using the notation of the lemma,  we set
$\tilde \beta_n:=\log \beta_n$,  $\tilde \beta_n^{(0)}:=\log \beta_n^{(0)}$,
and define $\tilde S_n^{(0)}:=(\xi_n^{(0)}, \tilde \beta_n^{(0)})$. Then
to given IF $\psi$  by the chain rule and \eqref{Lbtransf},
$\eta(x;\tilde\theta):= d_\beta^{-1} \psi(x;\theta)$ becomes an IF
in the log-scale model. By construction~\eqref{onestep},
$\tilde \beta_n = \tilde {\beta}_n^{(0)} + \frac{1}{n}\sum_i\eta_2(X_i;\tilde S_n^{(0)})$, so
\begin{align*}
\beta_n &= \beta_n^{(0)} \exp\left(\frac{1}{n}\sum_i\eta_2(X_i;\tilde S_n^{(0)})\right)=
\beta_n^{(0)} \exp\left(\frac{1}{{n\beta_n^{(0)}}}\sum_i \psi_2(X_i;S_n^{(0)})\right)
\end{align*}
So $\beta_n>0$ whenever $\beta_n^{(0)}$ is.
In particular, if $\sup_x |\psi_2(x;S_n^{(0)})|=b<\infty$,
the $\exp$-term remains in $[\exp(-b),\exp(b)]$, and hence breakdown (including implosion breakdown)
can occur iff breakdown has occurred in $\beta_n^{(0)}$.
\hfill
            \QEDlogo%\end{proof}
\medskip

\mparagraph{Proof of Lemma~\ref{Lem2.1}: }
We first note that $\alpha_0<x_0$, the positive zero of
$x\mapsto \log(1-x)+x+x^2$ (i.e., $x_0\doteq 0.6837$).
By the asymptotic linearity of MLE, if we use a suitable (uniformly integrable)
initialization, the bias of SMLE has the asymptotic representation
\begin{eqnarray}\label{BiasAppr}
B_n&=&n_\beta(\Ew ({\rm SMLE}) -\theta) = %n_\beta( \frac{1}{n}
%\sum_{k=1}^{\alpha_nn} \Ew \IF_{(\xi,\beta)}(z(X_{(n+1-k:n)}));{\rm MLE},F_\theta))=%\nonumber\\
%&=&
\Big((\frac{1}{n} |\sum_{k=1}^{\lceil \alpha_nn \rceil} \Ew \tilde \psi_\xi(V_{(k:n)})|^2
+ (\frac{1}{n} |\sum_{k=1}^{\lceil\alpha_nn\rceil}\Ew \tilde \psi_\beta(V_{(k:n)})|^2/\beta^2\Big)^{1/2}
\end{eqnarray}
for $X_{(k:n)}$, $V_{(k:n)}$ the respective $k$th order statistic.
Using \eqref{tpsidef}, we see that for $v\in(0,1)$,
the components of the $\IF$ of MLE  may each be written as $a \log(v) + f(v)$,
$a \not=0$, and $f$ bounded on this range. Hence the dominating term
is $\log(v)$.\\ As the order statistics $V_{(k:n)}$
are Beta-distributed,  we thus have to consider $|\Ew\log(B_{k,n})|$
for $B_{k,n}\sim {\rm Beta}(k,n-k+1)$, 
$k=1,\ldots,\lceil\alpha_n n\rceil$. To this end, note that by the power series
expansion of $\log(1-x)$, for any $L>0$ and any $x\in(0,1]$,
$
-\log(x) \ge \sum_{l=1}^L (1-x)^l /l
$,
while for $0\leq x<x_0$,
$
\log(1-x) \ge -x -x^2
$.
As
$1-B_{k,n}\sim {\rm Beta}(n-k+1,k)$, we further observe for $n>k$ that
$\Ew (1-B_{k,n})^l = \prod_{j=1}^l (n+j-k)/(n+j)$,
and that for any decreasing suitably integrable function $f(x)$
with (indefinite) integral $F(x)$,
$\sum_{j=1}^n f(j) \leq \int_{0}^n f(x) \,dx =F(n)-F(0)$.
Hence, using $1-x\leq e^{-x}$ for $x\in\R$ we obtain
\begin{eqnarray*}
&&E_{k,n}:=|\Ew\log(B_{k,n})| \ge
\sum_{l=1}^L \Ew (1-B_{k,n})^l /l \ge \sum_{l=1}^L
\frac{1}l \prod_{j=1}^l \Tfrac{n+j-k}{n+j}=%\\&=&
\sum_{l=1}^L \Tfrac{1}l \exp\big( \sum_{j=1}^l
\log (1- \Tfrac{k}{n+j})\big) \ge\\
&\ge&
\sum_{l=1}^L \Tfrac{1}l \exp\big( - \sum\limits_{j=1}^l
\Tfrac{k}{n+j}+\Tfrac{k^2}{(n+j)^2}\big) \ge%\\&\ge&
\sum\limits_{l=1}^L \Tfrac{1}{l} \exp\big( - k \log(\Tfrac{n+l}{n})
-\Tfrac{k^2l}{(n+l)n}\big) =\\
&=&
\sum\limits_{l=1}^L \Tfrac{1}{l} (1-\Tfrac{l}{n+l})^k
\exp(-\Tfrac{k^2l}{(n+l)n})\ge%\\
%&\ge&
\sum\limits_{l=1}^L \Tfrac{1}{l} (1-\Tfrac{L}{n+L})^k
\exp(-\Tfrac{k^2L}{(n+L)n})\ge %\\
%&\ge&
\log(L)(1-\Tfrac{L}{n+L})^k \exp(-\Tfrac{k^2L}{(n+L)n})
\end{eqnarray*}
Plugging in $L=\lceil\frac{1}{\alpha_n}\rceil$, we obtain,
eventually in $n$,
$E_{k,n}\ge -\log(\alpha_n) \exp(-1-\alpha_n) $.
On the other hand, for  $\beta_{1,n}$ the densitiy of ${\rm Beta}(1,n)$, we split
the integration range into $[0,1/n]$ and $[1/n,1]$ and obtain
\begin{eqnarray*}
0 &<&  \int_0^1 -\log(x)\,\beta_{1,n}(x)\,dx \leq n (\log(n)+1)/n + \log(n) \leq 3\log(n)
\end{eqnarray*}
if $n>2$. Now, for some constants $d_1,d_2\ge 0$ independent of $k$ and $n$,
$$%\begin{eqnarray*}
|\Ew \tilde\psi_\xi(B_{k,n})| =
\Tfrac{(\xi +1)^2}{\xi}E_{k,n} +
d_1-\Tfrac{\xi^2+3\xi+1}{\xi^2+\xi},\qquad
|\Ew \tilde\psi_\beta(B_{k,n})| =
\Tfrac{(\xi +1)}{\xi}E_{k,n} + d_2-(3-\Tfrac{1}{\xi})
$$%\end{eqnarray*}
Hence, as $\frac{\xi^2+3\xi+1}{\xi^2+\xi}< 3+\xi^{-1}$,
for $\liminf \alpha_n<\alpha_0$ we obtain, eventually in $n$
\begin{eqnarray*}
0&\le & \Tfrac{(\xi+1)\sqrt{(\xi+1)^2+\beta^{-2}}}{\xi} \alpha_n
(-\log(\alpha_n/\alpha_0)) \exp(-1-\alpha_n) \leq\\
&\leq&
\frac{1}{n}\sum_{k=1}^{\lceil\alpha_nn\rceil} \frac{\xi+1}{\xi}
\sqrt{((\xi+1)^2+\beta^{-2})\big( E_{k,n}-3-1/\xi\big)^2} \leq\\
&\leq&  \Big(\{\frac{1}{n}\sum_{k=1}^{\lceil\alpha_nn\rceil}
  \Ew \tilde\psi_\xi(B_{k,n})\}^2+\{\frac{1}{n}\sum_{k=1}^{\lceil\alpha_nn\rceil}
  \Ew \tilde\psi_\beta(B_{k,n})\}^2/\beta^2\Big)^{1/2}=B_n
\end{eqnarray*}
and $\liminf B_n>0$ if $\liminf \alpha_n>0$, respectively $\liminf n^\zeta B_n> c n^\zeta \alpha_n \log(n) $ if $\liminf n^\zeta\alpha_n>0$.
On the other hand, eventually in $n$ (as the other summand terms of
$\tilde \psi$ are bounded in $n$)
$$
B_n\leq  4 \frac{(\xi+1)\sqrt{(\xi+1)^2+1/\beta^2}}{\xi^2} \alpha_n \log(n)
$$
\hfill
            \QEDlogo%\end{proof}
\subsection*{Proofs of the Propositions in the Appendix}
\mparagraph{Proof of Proposition~\ref{MLEprop} (MLE): }
\begin{enumerate}
\item[{\textrm{\bf IF}}] The IF of MLE in our context has already been obtained in various references, see e.g.\ \citet{Smith:87};
as usual, we have $\IF_\theta(x;{\rm MLE},F_\theta) =  \mathcal{I}_\theta^{-1} \Lambda_\theta(x)$.
We have recalled the exact terms in \eqref{tpsidef} for later reference. Regularity conditions, e.g.  \citet[Thm.~5.39]{vdW:98}, can easily be
checked due to the smoothness of the scores function and entail that MLE attains the
smallest asymptotic variance among all ALEs according to the Asymptotic Minimax Theorem,
\citet[Thm.~3.3.8]{Ried:94}.
\item[{\textrm{\bf asVar}}] Again, the asymptotic covariance of MLE for its use in the Cram\'er Rao bound
has already been spelt out in other places, see e.g.\ \citep{Smith:87}.
\item[{\textrm{\bf asBias}}] As $({\cal I}_\theta^{-1})_{1,1},
    ({\cal I}_\theta^{-1})_{2,1}\not=0$,
both components of the joint IF are unbounded; the growth rate follows from \eqref{tpsidef}.
\item[{\textrm{\bf FSBP}}]
The assertion on FSBP follows easily by letting one observation tend to $\infty$.
Admittedly, for an actual finite sample,
one only can approximate this breakdown with extremely large contaminations.\hfill\QEDlogo\smallskip
\end{enumerate}

\mparagraph{Proof of Proposition~\ref{SMLEprop} (SMLE): }
\begin{enumerate}
\item[{\textrm{\bf IF}}] In fact, we follow the derivation of IFs to L-estimators in \citet[Ch. 3.3]{Hu:81}.
Up to bias $B_n$ we are interested in the $\alpha$-trimmed mean of the scores,
to which corresponds the functional given in  \eqref{SMLEfunc}.
Using the underlying order statistics of the $X_i$,
 we obtain \eqref{SMLE1} and \eqref{SMLE2}
as in the cited reference.
\item[{\textrm{\bf asVar}}] As $B_\theta$ is not random, the assertion is evident.
\item[{\textrm{\bf asBias}}] The assertion on the size of the bias follows from Lemma~\ref{Lem2.1}.
As the $\IF$ is bounded locally uniform in $\theta$, indeed the extra bias induced
by contamination is dominated by $B_n$ eventually in $n$.
\item[{\textrm{\bf FSBP}}]
In our shrinking setting the proportion of the skipped data
tends to  $0$, so it is the proportion which delivers the
active bound for the breakdown point: just replace
$\lceil\alpha_n n\rceil +1 $ observations by something sufficiently
large and argue as for the MLE to show that FSBP=$\alpha_n$. \hfill\QEDlogo\smallskip
\end{enumerate}

%\pagebreak
\mparagraph{Proof of Proposition~\ref{MDEprop} (MDE): }
\begin{enumerate}
\item[{\textrm{\bf IF}}] We follow
\citet[Example~4.2.15, Thm.~6.3.8]{Ried:94} and obtain
$
\IF(x;{\rm MDE},F_\theta) =:\mathcal{J_\theta}^{-1} (\tilde\varphi_\xi(x), \tilde\varphi_\beta(x))
$
with $\tilde\varphi$ as in the proposition
 and  $\mathcal{J}_\theta$ the
CvM Fisher information  as defined, e.g.\, in
\citet[Definition~2.3.11]{Ried:94}, i.e.;
$$
{\mathcal{J}_\theta}^{-1} =
3(\xi+3)^2\left(\begin{array}{cc}
\Tfrac{18(\xi+3)}{(2\xi+9)}, & -3\beta \\
-3\beta, & 2\beta^2
\end{array}\right)
$$
\item[{\textrm{\bf asVar}}] The asymptotic covariance of the CvM minimum distance estimators can be found
analytically or numerically. Our analytic terms are cross-checked against numeric evaluations;
{\tt MAPLE} scripts are available upon request for the interested reader.
\item[{\textrm{\bf asBias}}] The fact that the IF is bounded follows e.g.\ from
\citet[Example~4.2.15, 4.2 eq.(55), Thm.~6.3.8, Rem~6.3.9(a)]{Ried:94}.
\item[{\textrm{\bf FSBP}}]
Due to the lack of invariance in the GPD situation,
\citet[Propositions~4.1 and~6.4]{Do:Li:88}
only provide lower bounds for the FSBP, which is $1/2$ the FSBP of the
FSBP-optimal procedure among  all Fisher con\-sistent estimators.

As ${\rm MDE}$ is a minimum of the smooth CvM distance,
it has to fulfill the first order condition for
the corresponding M-equation, i.e.; for
$V_i=(1+\frac{\xi}{\beta}X_i)^{-1/\xi}$,
$$\sum\nolimits_i \tilde\varphi_\xi(V_i;\xi) = 0,\qquad
\sum\nolimits_i \tilde\varphi_\beta(V_i;\xi) = 0
$$
Arguing as for the breakdown point of an M-estimator,
except for the optimization in $\xi$, we obtain \eqref{BDPbound} as an
 analogue to \citet[Ch.~3, eqs.~(2.39) and (2.40)]{Hu:81}.

In our shrinking setting the proportion of the skipped data
tends to  $0$, so it is the proportion which delivers the
active bound for the breakdown point: just replace
$\lceil\alpha_n n\rceil +1 $ observations by something sufficiently
large and argue as for the MLE to show that FSBP=$\alpha_n$. \hfill\QEDlogo\smallskip
\end{enumerate}

\mparagraph{Proof of Proposition~\ref{PEprop} (PE): }
\begin{enumerate}
\item[{\textrm{\bf IF}}] The IF of linear combinations $T_L$ of
the quantile functionals $F^{-1}(\alpha_i) = T_i(F)$
for probabilities $\alpha_i$ and weights $h_i, \ i = 1,...,k$
may be taken from \citet[Ch. 1.5]{Ried:94} and gives
$$%\begin{equation}
\IF(x;T_L,F_\theta) = \sum\nolimits_{i = 1}^{k} h_i\,
\big(\alpha_i - \mathbb{I}(x \leq F^{-1}(\alpha_i))\big)/f(F^{-1}(\alpha_i))
$$%\end{equation}
Using the $\Delta$-method, the IFs of PE(a) hence is
$$%\begin{align}
\IF_\Bullet (x;{\rm PE}(a),F_\theta) =   \sum\nolimits_{i = 2,3} h_{\Bullet,i}(a)
\Tfrac{\alpha_i(a) - \mathbb{I}(x \leq \hat Q_{i}(a))}{f(\hat Q_{i}(a))}, \qquad \Bullet = \xi,\beta %\\
$$%\end{align}
with weights $h_{\Bullet,i}(a)$ which for $\hat Q_{i}=\hat Q_{i}(a)$, $i=2,3$ are given by
\begin{align*}
h_{\xi,2}(a) & = -\frac{1}{\log(a)} \frac{\hat Q_3}{\hat Q_2 (\hat Q_3 - \hat Q_2)}, \\
h_{\beta,2}(a) &= h_{\xi,2}(a)\, \frac{(\hat Q_2)^2}{\hat Q_3 - 2 \hat Q_2}
+ \frac{1}{\log(a)} \frac{2\hat Q_2(\hat Q_3-\hat Q_2)}{(\hat Q_3-2\hat Q_2)^2} \,\log \frac{\hat Q_3 - \hat Q_2}{\hat Q_2}
\\
h_{\xi,3}(a) &= \frac{1}{\log(a)} \frac{1}{\hat Q_3- \hat Q_2},\\
h_{\beta,3}(a) &= h_{\xi,3}(a) \,\frac{(\hat Q_2)^2}{\hat Q_3 - 2 \hat Q_2}
- \frac{1}{\log(a)}\frac{(\hat Q_2)^2}{(\hat Q_3-2\hat Q_2)^2}\,\log \frac{\hat Q_3 - \hat Q_2}{\hat Q_2}
\end{align*}
\item[{\textrm{\bf asVar}}] This follows from integrating out the IF.
\item[{\textrm{\bf asBias}}] Boundedness of the IF is obvious from the terms just derived, so {\rm asBias} is finite.
\item[{\textrm{\bf FSBP}}]
Terms for $\ve^\ast_n $ are simple generalizations of \citet[Prop.~5.1]{Ru:Ho:10b},
 $\bar \ve^\ast$ follows from usual LLN arguments.\hfill\QEDlogo\smallskip
\end{enumerate}

\mparagraph{Proof of Proposition~\ref{MMedprop} (MMed): }
A general reference is  \citet{P:W:01}.
\begin{enumerate}
\item[{\textrm{\bf IF}}] The IF of MMed is a linear combination of the IF
of the sample median already used for the PE, and the IF of
the median of the $\xi$-coordinate of $\Lambda_{\theta_1;2}(X)$.
The assertion on the level sets of form $[q_1,q_2]$ follows from \citet{P:W:01}
or by plotting the respective IF for actual $\xi$-values.
More precisely, for $\xi=0.7$ we obtain $q_1\doteq 0.3457$ and
$q_2\doteq 2.5449$.\\
\eqref{lmmedA} is a simple generalization of the IF to a general
quantile and \eqref{tildeDdef} is entailed by the $\Delta$-method.
As $D$ does not depend on $x$, we may incorporate the standardizing
term involving evaluations of $f_\theta$ into $\tilde D$ and to obtain the IF
as $%\begin{equation}
{\rm IF}(x;{\rm MMed},F_\theta)=\tilde D\chi_\theta
$ with $\chi_\theta$ from \eqref{tildeDdef}.  %\end{equation}
\item[{\textrm{\bf asVar}}] This follows from integrating out the IF.
\item[{\textrm{\bf asBias}}] The IF of MMed is clearly bounded, so {\rm asBias} is finite.\hfill\QEDlogo\smallskip
\end{enumerate}

\mparagraph{Proof of Proposition~\ref{MedkMADprop} (MedkMAD): }
\begin{enumerate}
\item[{\textrm{\bf IF}}] By the implicit function theorem, the Jacobian in the Delta method is $D$ from
\eqref{DdefMedkMAD}. Hence by the $\Delta$-method, $%\begin{align}
\IF(x;{\rm MedkMAD},F_\theta) = D\,  (\IF(x;{\rm kMAD},F_\theta),\IF(x;{\rm median},F_\theta)l)^\tau
$ %\end{align}
where the IF of kMAD is a simple
generalization of the one for MAD, to be drawn e.g.\ from \citet[Ch. 1.5]{Ried:94}. For the entries of $D$  we note
\begin{align*}
\Tfrac{\partial G^{(1)}}{\partial \xi} &=
- v\left(\Tfrac{v^{\xi}-1}{\xi^2} -\Tfrac{1}{\xi}\log(v) \right)\,%
\Big|_{v=v_-}^{v_+}, \quad\,\nonumber
\Tfrac{\partial G^{(1)}}{\partial \beta} =
 \Tfrac{v}{\xi\beta^2}(v^\xi-1)\,\Big|_{v=v_-}^{v_+}, \quad \\
\Tfrac{\partial G^{(2)}}{\partial \xi} &= \Tfrac{\beta}{\xi}
\left(2^\xi \log(2) - \Tfrac{2^\xi-1}{\xi} \right), \quad %\nonumber\\
\Tfrac{\partial G^{(2)}}{\partial \beta} = \Tfrac{2^\xi-1}{\xi}, \quad \nonumber\\
\Tfrac{\partial G^{(1)}}{\partial M} &=
\Tfrac{k v_+^{\xi+1}+v_-^{\xi+1}}{\beta},\quad
\Tfrac{\partial G^{(1)}}{\partial m} =
\Tfrac{v^{\xi+1}}{\beta}\,\Big|_{v=v_-}^{v_+},\;\;
\Tfrac{\partial G^{(2)}}{\partial M} = 0, \quad
\Tfrac{\partial G^{(2)}}{\partial m} = -1
\end{align*}
for 
\begin{equation}
v_+:=\left(1+\xi\Tfrac{kM+m}{\beta} \right)^{-\frac{1}{\xi}},\qquad
v_-:=\left(1+\xi\Tfrac{m-M}{\beta} \right)^{-\frac{1}{\xi}} \nonumber
\end{equation}

\item[{\textrm{\bf asVar}}] With obvious generalizations, $\sigma_{i,j}$, $i,j=1,2$, may be taken from
\citet{S:M:09}.
\item[{\textrm{\bf asBias}}]
Both IFs of median and kMAD are bounded, so the asymptotic bias of MedkMAD is finite.
\item[{\textrm{\bf FSBP}}]
The assertions are shown in  \citet[Prop.~5.2]{Ru:Ho:10b}.\hfill\QEDlogo\smallskip
\end{enumerate}

\normalsize
\section*{Acknowledgement}
We thank two anonymous referees for their valuable and helpful comments.

\bibliographystyle{spmpsci}
\begin{footnotesize}

\end{footnotesize}

\label{lastpage}
\end{document}